\documentclass[aps,pre,twocolumn,amsmath,amssymb,superscriptaddress,showpacs,longbibliography]{revtex4-1}

\usepackage{graphicx}
\usepackage{dcolumn}
\usepackage{bm}
\usepackage{mathrsfs}

\usepackage{natbib}
\usepackage[text={7in,9.5in},centering]{geometry}

\usepackage{hyperref}
\usepackage{url}

\usepackage{epsfig}
\usepackage{amsmath}
\usepackage{amssymb}
\usepackage{textcomp}

\begin{document}

\title{Nonbacktracking expansion of finite graphs}

\author{G. Tim\'ar}
 \affiliation{Departamento de F\'\i sica da Universidade de Aveiro \& I3N, Campus Universit\'ario de Santiago, 3810-193 Aveiro, Portugal}

\author{R. A. da Costa}
 \affiliation{Departamento de F\'\i sica da Universidade de Aveiro \& I3N, Campus Universit\'ario de Santiago, 3810-193 Aveiro, Portugal}

\author{S. N. Dorogovtsev}
 \affiliation{Departamento de F\'\i sica da Universidade de Aveiro \& I3N, Campus Universit\'ario de Santiago, 3810-193 Aveiro, Portugal}
 \affiliation{A. F. Ioffe Physico-Technical Institute, 194021 St. Petersburg, Russia}

\author{J. F. F. Mendes}
 \affiliation{Departamento de F\'\i sica da Universidade de Aveiro \& I3N, Campus Universit\'ario de Santiago, 3810-193 Aveiro, Portugal}


\begin{abstract}
Message passing equations yield a sharp percolation transition in finite graphs, as an artifact of the locally treelike approximation.
For an arbitrary finite, connected, undirected graph we construct an infinite tree having the same local structural properties as this finite graph, when observed by a nonbacktracking walker. Formally excluding the boundary, this infinite tree is a generalization of the Bethe lattice.
We indicate an infinite, locally treelike, random network whose local structure is exactly given by this infinite tree.
Message passing equations for various cooperative models on this construction are the same as for the original finite graph, but here they provide the exact solutions of the corresponding cooperative problems. 
These solutions are good approximations to observables for the models on the original graph when it is sufficiently large and not strongly correlated.
We show how to express these solutions in the critical region in terms of the principal eigenvector components of the nonbacktracking matrix. 
As representative examples we 
formulate the problems of the random and optimal destruction of a connected graph in terms of our construction, the nonbacktracking expansion. 
We analyze the limitations and the accuracy of the message passing algorithms for different classes of networks and compare the complexity of the message passing calculations to that of direct numerical simulations. Notably, in a range of important cases, simulations turn out to be more efficient computationally than the message passing. 
\end{abstract}

\maketitle


\section{Introduction}
\label{s1}

The Bethe lattice is a valuable substrate, on top of which many cooperative models 
are solved exactly \cite{baxter2007exactly} and show critical behaviors with mean-field critical exponents and Gaussian critical fluctuations. 
A regular Bethe lattice is equivalent to an infinite random regular graph, which is locally treelike and has only infinite cycles (loops) and no boundary. This is in contrast with an infinite Cayley tree in which the boundary 
suppresses phase transitions, e.g., in the Ising model on a Cayley tree, finite or infinite, long-range order never emerges. The configuration model \cite{bender1978asymptotic,bollobas1980probabilistic} of a sparse uncorrelated network and its generalizations represent a heterogeneous version of the Bethe lattice.  This basic model also provides exact solutions 
for percolation and other problems in infinite random networks \cite{newman2001random}.

Self-consistency equations for the probability of reaching a finite branch by following a link yield the solution of percolation problems for infinite locally treelike random networks \cite{newman2001random}. 
Recently it was found in Ref.~\cite{karrer2014percolation} that such self-consistency equations 
produce a good approximation of the percolation behavior of large but still finite, sparse graphs. This  conclusion was supported by numerical simulations. 
The equations are written for a given graph 
from which a fraction $1-p$ of links is removed. 
These self-consistency equations---termed message passing equations in the context of finite graphs---predict a sharp continuous phase transition for any finite graph, e.g., for the $4$-clique (four interconnected nodes). 
In reality, a percolation continuous phase transition cannot be defined in finite systems. The same is true for other models on finite graphs.
In this paper, instead of describing a given finite system with some unknown accuracy, we address the issue of this useful approximation from a complementary perspective.
Namely, if the message passing equations in principle cannot precisely describe physical models on finite graphs, then what do they describe?
For each given finite graph, we construct a related infinite network for which these equations provide an exact description of the cooperative models.
This infinite network is a generalization of the Bethe lattice with percolation properties that are described exactly by the message passing equations on the finite graph \cite{karrer2014percolation}.
The closer the original finite graph is to this construction, the better the approximation provided by the message passing equations.    
We therefore give a physical interpretation of the solutions of these equations for percolation on a finite graph.  
We express these solutions in the critical region in terms of the principal eigenvector components of the nonbacktracking matrix. 
As another example we consider the problem of optimal percolation \cite{morone2015influence,morone2016collective}. We show that using our network construction, an approximate algorithm for finding the so-called decycling number \cite{bau2002decycling} can be formulated in simple terms of the principal eigenvector of the nonbacktracking matrix. 
We introduce a way to quantify the accuracy of the message passing algorithm and indicate two classes of networks for which the approximation fails. 
We calculate the time complexity associated with message passing algorithms and show that in a wide range of important situations this method is less efficient than direct numerical simulations.


\section{Construction}
\label{s2}

\begin{figure}[t]
\centering
\includegraphics[width=6cm,angle=0.]{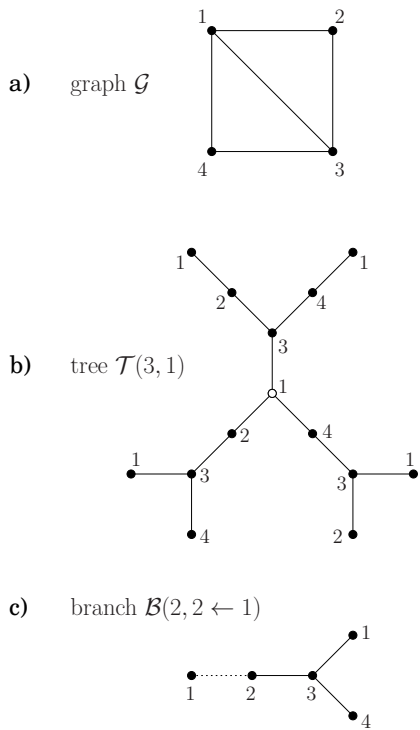}
\caption{Construction of (b) the tree $\mathcal{T}(3,1)$ and (c) the branch $\mathcal{B}(2, 2 \leftarrow 1)$ for an example graph (a). The root (starting node) in $\mathcal{T}(3,1)$ is shown by the open circle and the starting link in $\mathcal{B}(2, 2 \leftarrow 1)$ is represented by the dotted line.}
\label{fig:trees}
\end{figure}

To build the \textit{nonbacktracking expansion} (NBE) of an arbitrary finite, connected, undirected graph $\mathcal{G}$, we first construct the tree $\mathcal{T}(\ell,i)$ in the following way.
For any given node $i$ of $\mathcal{G}$,
$\mathcal{T}(1,i)$ is a star graph consisting of a node labeled $i$ in the center, and for any node $j$ in
$\mathcal{G}$ for which a link $j {\leftarrow} i$ exists, a node labeled $j$ is attached to the center node of $\mathcal{T}(1,i)$.
For general $\ell > 1$ we define $\mathcal{T}(\ell,i)$ in a recursive way.
To any leaf labeled $k$ in $\mathcal{T}(\ell-1,i)$, we attach a node labeled $l$ if and only if
$B_{l \leftarrow k, k \leftarrow j} = 1$, where $\mathbf{B}$ is the nonbacktracking matrix of graph $\mathcal{G}$ and $j$
is the label of the parent node of leaf $k$ in $\mathcal{T}(\ell-1,i)$.  The nonbacktracking matrix \cite{hashimoto2014automorphic, krzakala2013spectral} 
is defined as $B_{i \leftarrow j, k \leftarrow l} = \delta_{jk}(1-\delta_{il})$,
where $i,j$ and $k,l$ are end node pairs of links in the original graph. Note that both 
directions must be taken into account for each link.
We see that the branches emanating from the root
$i$ in $\mathcal{T}(\ell,i)$ correspond exactly to nonbacktracking walks of, at most, length $\ell$, in the original graph $\mathcal{G}$
starting from node $i$. (Note that some of these walks may be shorter than $\ell$, when a dead end in graph $\mathcal{G}$ is encountered.)
In this construction the nodes of the resulting tree have only labels of the original finite graph, $i=1,2,\ldots,N$, so different nodes in this tree can have the same labels. The same is true for nodes in each individual branch of the tree.
Figure~\ref{fig:trees}(b) explains $\mathcal{T}(\ell,i)$ for the finite graph in Fig.~\ref{fig:trees}(a).
The construction $\mathcal{T}(\ell,i)$ is also known as a computation tree in the computer science community \cite{weiss2001optimality}.
Taking the limit $\ell\to\infty$ and formally excluding the boundary, we obtain a network in which all nodes of the same label $i$ and all links of the same label $j {\leftarrow} i$ are topologically equivalent, in the sense that the structure of any finite neighborhood of each node of label $i$ (and of each link of label $j {\leftarrow} i$) is the same.
This construction is a generalization of the Bethe lattice and we call it the nonbacktracking expansion of graph $\mathcal{G}$. It is a specific Bethe lattice 
preserving the local topology of connections of the graph $\mathcal{G}$. For example, the NBE of the $4$-clique 
is simply a regular Bethe lattice of coordination number 3.

\begin{figure}[htpb!]
\centering
\includegraphics[width=6cm,angle=0.]{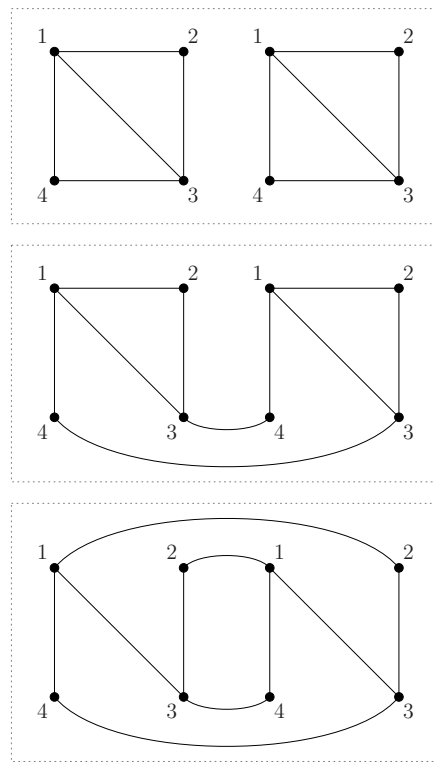}
\caption{Construction of network $\mathcal{N}(2)$ for the finite graph in Fig.~\ref{fig:trees}(a), according to Ref.~\cite{faqeeh2015network}. Three of the members of the statistical ensemble of graphs constituting $\mathcal{N}(2)$ are shown as examples.}
\label{fig:bethe}
\end{figure}

A more explicit definition of the NBE can be given using the ``$m$-cloned'' network \cite{faqeeh2015network} of the given graph $\mathcal{G}$.
This network is constructed as follows.
Let the nodes of $\mathcal{G}$ be labeled $i=1,2,\ldots,N$. We make $m$ 
replicas of all these nodes, still labeled $i=1,2,\ldots,N$ within each of the $m$ replicas. The $mN$ nodes obtained
in this way are the nodes of the $m$-cloned network $\mathcal{N}(m)$ defined as follows.
Let $\mathcal{N}(m)$ be the network that
is maximally random under the constraint that if and only if nodes $i$ and $j$ are connected in $\mathcal{G}$, then any of
the $m$ nodes labeled $i$ in $\mathcal{N}(m)$ is connected to exactly one node
labeled $j$ (of, in total, $m$ such nodes). In other words, make $m$ copies of graph $\mathcal{G}$ and consider all possible rewirings that preserve the labels of end nodes of all links, with equal statistical weights (uniform randomness).  
Figure~\ref{fig:bethe} explains this construction for $m=2$.
Let us take the limit $m\to\infty$ and consider the infinite network $\mathcal{N}(m \to\infty)$.
This network is locally treelike: any finite neighborhood of any node labeled $i$ is given by the tree $\mathcal{T}(\ell,i)$.
The relative number of finite loops in this infinite network is negligible, almost all loops are infinite, and almost all nodes of label $i$ (and links of label $j {\leftarrow} i$) are topologically equivalent.
The network $\mathcal{N}(m \to\infty)$ gives exactly the NBE of graph $\mathcal{G}$.

To discuss the properties of the NBE it is helpful to
define the branch $\mathcal{B}(\ell, j \leftarrow i)$ as the tree that is recursively generated by the nonbacktracking walks of length
$\ell$ in $\mathcal{G}$ starting from the link $j {\leftarrow} i$, as shown in Fig.~\ref{fig:trees}(c). 
Clearly, $\mathcal{T}(\ell+1,i)$ consists of all branches $\mathcal{B}(\ell, j \leftarrow i)$ (all $j$, for
which there is a link $j \leftarrow i$ in graph $\mathcal{G}$).


\section{Properties of the nonbacktracking expansion}
\label{s3}

Let us consider 
properties of these constructions.
Let $n_{l \leftarrow k}(\ell, j \leftarrow i)$ be the number of links of label $l \leftarrow k$ 
on the surface of $\mathcal{B}(\ell, j \leftarrow i)$ (these are links from parent $k$ on the surface of sphere $\ell-1$ to
child $l$ on the surface of sphere $\ell$). Let $n(\ell, j \leftarrow i)$ be the size (number of nodes) of the
surface of $\mathcal{B}(\ell, j \leftarrow i)$, so that 
$f_{l \leftarrow k}(\ell, j \leftarrow i) \equiv n_{l \leftarrow k}(\ell, j \leftarrow i) / n(\ell, j \leftarrow i)$ is the relative
number of links of label $l \leftarrow k$ on the surface of $\mathcal{B}(\ell, j \leftarrow i)$.
Every link on the surface of $\mathcal{B}(\ell+1, j \leftarrow i)$ must emanate from the end node of a link on the
surface of $\mathcal{B}(\ell, j \leftarrow i)$, so 
\begin{equation}
   n_{l \leftarrow k}(\ell+1, j \leftarrow i) = \sum_{l' \leftarrow k'} n_{l' \leftarrow k'}(\ell, j \leftarrow i)
   B_{l \leftarrow k, l' \leftarrow k'}
   ,
  \label{eq:10}
\end{equation}
that is,
\begin{multline}
   n(\ell+1, j \leftarrow i) f_{l \leftarrow k}(\ell+1, j \leftarrow i)  \\
   = n(\ell, j \leftarrow i) \sum_{l' \leftarrow k'} f_{l' \leftarrow k'}(\ell, j \leftarrow i) B_{l \leftarrow k, l' \leftarrow k'}
   .
  \label{eq:20}
\end{multline}
Taking $\ell \rightarrow \infty$, the ratio $n(\ell+1, j \leftarrow i) / n(\ell, j \leftarrow i)$ 
converges to an asymptotic
mean branching $\overline{b}$ and the numbers $f_{l \leftarrow k}(\ell, j \leftarrow i)$ must also converge to some
asymptotic values $f_{l \leftarrow k}^*(j \leftarrow i)$. For $\ell \rightarrow \infty$, therefore, we 
get
\begin{equation}
   \overline{b} f_{l \leftarrow k}^*(j \leftarrow i) = \sum_{l' \leftarrow k'} f_{l' \leftarrow k'}^*(j \leftarrow i)
   B_{l \leftarrow k, l' \leftarrow k'}
  \label{eq:30}
\end{equation}
or, in vector form, 
\begin{equation}
   \overline{b} \mathbf{f}^*(j \leftarrow i) = \mathbf{B} \mathbf{f}^*(j \leftarrow i)
   .
  \label{eq:40}
\end{equation}
Equation (\ref{eq:40}) is 
a right eigenvector equation for the nonbacktracking matrix of the original graph $\mathcal{G}$. According to the
Perron-Frobenius theorem \cite{minc1988nonnegative}, 
the nonbacktracking matrix $\mathbf{B}$ has exactly one strictly non-negative right eigenvector, and this is the eigenvector
corresponding to the leading eigenvalue. All of the components of $\mathbf{f}^*(j \leftarrow i)$ are, by definition,
non-negative, therefore the mean branching $\overline{b}$ is equal to the leading eigenvalue $\lambda_1$ of $\mathbf{B}$.
We note that the leading eigenvalue and the corresponding right eigenvector are independent of the starting link $j \leftarrow i$,
so we have the following general result. The asymptotic mean branching of $\mathcal{B}(\ell, j \leftarrow i)$, for any
$j \leftarrow i$, is given by the leading eigenvalue $\lambda_1$ of the nonbacktracking matrix of graph $\mathcal{G}$.
From the construction of $\mathcal{T}(\ell, i)$, it is clear that $\mathcal{T}(\ell, i)$ must also have the
same asymptotic mean branching for any 
root $i$.
The asymptotic relative numbers of links $l \leftarrow k$ on the surface of $\mathcal{B}(\ell, j \leftarrow i)$
[or $\mathcal{T}(\ell, i)$] are given by the components of the principal right eigenvector $\mathbf{v}^{(\lambda_1)}$ of $\mathbf{B}$. These asymptotic relative numbers are, again, independent of the starting link
(or node).

Now we can 
find the asymptotic relative number $g_i$ of nodes of label $i$ on the surface [of either
$\mathcal{B}(\ell, r' \leftarrow r)$ or $\mathcal{T}(\ell, r)$, independent of the starting link $r' \leftarrow r$ or root $r$]: 
\begin{equation}
   g_i = \frac{\sum_{j \in 
   \partial i} v_{i \leftarrow j}^{(\lambda_1)} }{ \sum_{k=1}^N \sum_{j \in 
   \partial k} v_{k \leftarrow j}^{(\lambda_1)} }
   ,
  \label{eq:50}
\end{equation}
where $
 \partial i$ 
 denotes the set of nearest neighbors of node $i$ in graph $\mathcal{G}$.
The asymptotic relative number of nodes $i$ must also be the same on the layer just beneath the surface, and each node in
this layer (apart from dead ends) must be the starting node of a link on the surface, so we can also write
\begin{equation}
   g_i = \frac{ \frac{1}{q_i - 1}  \sum_{j \in 
    \partial i} v_{j \leftarrow i}^{(\lambda_1)} }{ \sum_{k=1}^N  (\frac{1}{q_k - 1}  \sum_{j \in 
     \partial k} v_{j \leftarrow k}^{(\lambda_1)} )}
   ,
  \label{eq:60}
\end{equation}
where $q_i > 1$ is the degree of nodes $i$. Equations~(\ref{eq:50}) and (\ref{eq:60}) give a general relationship between
the components of the eigenvector $\mathbf{v}^{(\lambda_1)}$. Note that $g_i$ is the asymptotic probability of finding a nonbacktracking random walker at node $i$ after an infinitely long walk.
This quantity coincides with the nonbacktracking centrality of \cite{martin2014localization}.

Let us 
return to $n(\ell, j \leftarrow i)$, the size of the surface of $\mathcal{B}(\ell, j \leftarrow i)$. Let
$n(\ell)\equiv \sum_{j \leftarrow i} n(\ell, j \leftarrow i)$ be the sum of the sizes of all such surfaces, 
and
$h_{j \leftarrow i}(\ell)\equiv n(\ell, j \leftarrow i) / n(\ell)$ be the relative size of the surface of $\mathcal{B}(\ell, j \leftarrow i)$. 
Then 
\begin{equation}
   n(\ell+1, j \leftarrow i) = \sum_{l \leftarrow k} n(\ell, l \leftarrow k) B_{l \leftarrow k, j \leftarrow i}
  \label{eq:70}
\end{equation}
or
\begin{equation}
   n(\ell+1) h_{j \leftarrow i}(\ell+1) = n(\ell) \sum_{l \leftarrow k} h_{l \leftarrow k}(\ell) B_{l \leftarrow k, j \leftarrow i}
   .
  \label{eq:80}
\end{equation}
In the limit $\ell \rightarrow \infty$, the ratio $n(\ell+1) / n(\ell)$  
converges to the asymptotic mean
branching $\overline{b}$ and the numbers $h_{j \leftarrow i}(\ell)$ must also converge to some asymptotic values
$h_{j \leftarrow i}^*$.
We can write
\begin{equation}
   \overline{b} h_{j \leftarrow i}^* = \sum_{l \leftarrow k} h_{l \leftarrow k}^* B_{l \leftarrow k, j \leftarrow i}
  \label{eq:90}
\end{equation}
or, in vector form,
\begin{equation}
   \overline{b} \mathbf{h}^* = \mathbf{h}^* \mathbf{B}
   ,
  \label{eq:100}
\end{equation}
which is just a left eigenvector equation for $\mathbf{B}$.
The same reasoning applies as before, so $\overline{b}$ is equal to the leading eigenvalue of $\mathbf{B}$,
$\lambda_1$. The vector $\mathbf{h}^*$ is equal to the corresponding left eigenvector
$\mathbf{w}^{(\lambda_1)}$. 
Then the asymptotic relative sizes of the surfaces of
$\mathcal{B}(\ell, j \leftarrow i)$ are given by the components of $\mathbf{w}^{(\lambda_1)}$. We 
express the asymptotic relative sizes $s_i$ of the surfaces of $\mathcal{T}(\ell, i)$ in terms of these components:
\begin{equation}
   s_i = \frac{\sum_{j \in 
   \partial i} w_{j \leftarrow i}^{(\lambda_1)} }{ \sum_{k=1}^N \sum_{j \in 
   \partial k} w_{j \leftarrow k}^{(\lambda_1)} }
   .
  \label{eq:110}
\end{equation}
From the symmetric relationship between the left and the right eigenvectors
we see that
$w_{j \leftarrow i}^{(\lambda_1)} = v_{i \leftarrow j}^{(\lambda_1)}$, therefore
\begin{equation}
   s_i  = 
     g_i
   .
  \label{eq:120}
\end{equation}
Thus the asymptotic relative size of the surface of $\mathcal{T}(\ell, i)$ is equal to the asymptotic relative number of
nodes of label $i$ on the surface of any $\mathcal{T}(\ell, r)$ independently of $r$. This quantity 
is expressed in terms of the components of the principal (left or right) eigenvector
of the nonbacktracking matrix of the original graph $\mathcal{G}$.

\begin{figure}[t]
\centering
\includegraphics[width=7cm,angle=0.]{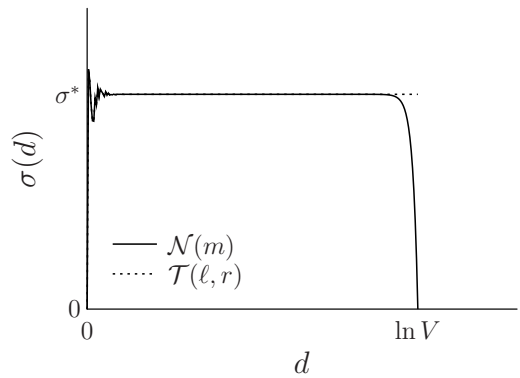}
\caption{
Dispersion of the relative number 
of node labels within a distance $d$ of an arbitrarily chosen node in the nonbacktracking expansion $\mathcal{N}(m)$ and the tree $\mathcal{T}(\ell, r)$. The number of nodes in both $\mathcal{N}(m)$ and $\mathcal{T}(\ell, r)$ is denoted by $V$.}
\label{fig:approach}
\end{figure}

It is clear from the above constructions that for an arbitrary graph $\mathcal{G}$, the interior of
the infinite network 
$\mathcal{N}(m\to\infty)$
is given exactly by 
$\mathcal{T}(\ell\to\infty, r)$,
independently of 
$r$. 
Note, however, that the distribution of the numbers of nodes of different labels in
$\mathcal{T}(\ell\to\infty, r)$ is given by Eq.~(\ref{eq:50}), while the same distribution
for 
$\mathcal{N}(m\to\infty)$ is uniform---we made the same $m$ number of copies of every node.
This discrepancy is reconciled by the presence of infinite loops: nodes in the infinite network $\mathcal{N}(m\to\infty)$ 
are ``matched up'' at infinity to provide the overall uniform distribution of node labels. 
It is informative to consider the 
dispersion $\sigma(d)$ of the 
relative number of node labels within a distance $d$ of an
arbitrarily chosen starting node in both $\mathcal{T}(\ell, r)$ and $\mathcal{N}(m)$. Suppose that $\ell$ and
$m$ are such that the number of nodes in both $\mathcal{T}(\ell, r)$ and $\mathcal{N}(m)$ is $V$ and that
$V$ is large. The dispersion $\sigma(d)$ in both cases will converge to
\begin{equation}
   \sigma^* = \langle g^2 \rangle - \langle g \rangle ^2
  \label{eq:130}
\end{equation}
but will go to $0$ for $\mathcal{N}(m)$ as $d$ approaches $\ln V$, while it remains the same for any $d$ in the
case of $\mathcal{T}(\ell, r)$. Here the variables $g$ are the asymptotic values given by Eq.~(\ref{eq:50}). 
Figure~\ref{fig:approach} qualitatively illustrates this behavior.

\section{Example Problems}
\label{s4}

We now consider the problem of approximating percolation on finite graphs. 
The critical singularity in percolation occurs only in infinite networks where the giant connected component can be strictly defined. In a large though finite random graph the so-called scaling window, in which the behavior of the largest connected cluster noticeably deviates from the singular behavior of the giant connected component, is narrow.  
In a recent work Karrer \textit{et al.} \cite{karrer2014percolation} considered this problem of ``approximate'' percolation on a finite graph by
solving self-consistency equations for the probabilities of reaching finite clusters by following a link in a given direction. Such
self-consistency equations provide an exact solution for percolation in infinite locally treelike random networks, and in finite graphs they
give an approximation to the behavior of the largest connected cluster. The larger and more ``locally treelike'' the finite graph is,
the better this approximation is expected to be. For an arbitrary connected finite graph the ``approximate critical point'' is given
by the inverse of the leading eigenvalue of the nonbacktracking matrix of the graph.
It was shown in Ref.~\cite{hamilton2014tight} that for infinite graphs, this is a tight lower bound on the exact percolation threshold.
The question arises: More than treating it purely as an approximation, what is the exact meaning of the solutions of self-consistency
equations on a finite graph?
The results in Ref.~\cite{karrer2014percolation} rely on solving the following set of basic equations:
\begin{equation}
  H_{i \leftarrow j}(x) = 1-p + px \prod_{k \in 
  \partial j \setminus i} H_{j \leftarrow k}(x)
  \label{eq:140}
  ,
\end{equation}
where $H_{i \leftarrow j}(x)$ is the probability generating function for the distribution of the sizes of finite clusters reachable
by following links $i \leftarrow j$ from $i$ to $j$. 
(In the context of finite graphs, the meaning of finite clusters is: not the
largest cluster.) 
Here $\partial j \setminus i$ denotes the set of the nearest neighbors of node $j$ excluding node $i$, and $p$ is the existence probability of any link.
Clearly, in finite graphs, these equations do not hold exactly, but may be approximately correct if the graph in
question is large and sparse. In our nonbacktracking expansion [the network $\mathcal{N}(m\to\infty)$],
however, Eqs.~(\ref{eq:140}) hold exactly, and by solving them we actually obtain the exact percolation results for
$ \mathcal{N}(m\to\infty)$. In Ref.~\cite{karrer2014percolation} it is shown that the value of $p$, at
which a non-trivial solution to Eqs.~(\ref{eq:140}) appears, is given by the inverse of the leading eigenvalue of the
nonbacktracking matrix of the given graph. This coincides with our result that the mean branching of the network
$\mathcal{N}(m\to\infty)$ is also given by the leading eigenvalue of the nonbacktracking matrix.
The inverse of this value is a true critical point in our NBE.
The order parameter is the relative size of the giant connected component, and is given by
\begin{equation}
  S = \frac{1}{N} \sum_i \left(  1-\prod_{j \in \partial i} H_{i \leftarrow j}(1) \right)  
  ,
  \label{eq:143}
\end{equation}
where $H_{i \leftarrow j}(1)$ are the solutions of Eq.(\ref{eq:140}) at $x=1$, for a given value of $p$.
The degree distribution of any NBE coincides with the degree distribution of the original finite graph, therefore it necessarily has a cutoff---the maximum degree in the original graph. Consequently the order parameter exponent $\beta = 1$, $S \cong \Omega (p-p_c)^\beta$, for the NBE of any finite graph.

\begin{figure}[t]
\centering
\includegraphics[width=7cm,angle=0.]{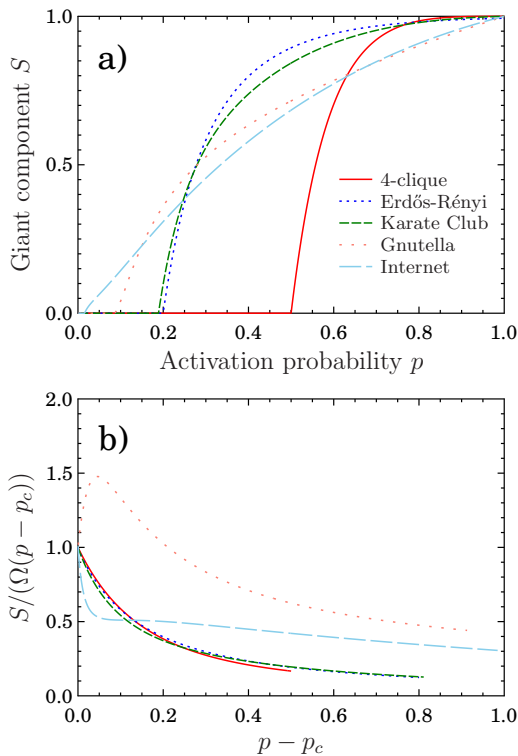}
\caption{(Color online) (a) The order parameter $S$ as a function of $p$, computed using Eq. (\ref{eq:143}), for the 4-clique, an Erd\H{o}s--R\'enyi graph, the Zachary Karate Club network \cite{zachary1977information}, a subgraph of the Gnutella peer-to-peer file sharing network \cite{network_gnutella}, and a snapshot of the Internet at the autonomous systems level \cite{meyer2001university}. (b) Normalized slopes of the same example networks, obtained using Eq. (\ref{eq:145}) [or Eq. (\ref{eq:145b})]. All normalized slopes go to $1$ at $p=p_c$.}
\label{fig:curves}
\end{figure}

We express the slope $\Omega$ of the order parameter at $p_c$ in terms of the components of the principal eigenvector of the nonbacktracking matrix which naturally emerges when linearizing Eq.~(\ref{eq:140}). We cannot do this directly, unlike, e.g.,  in Refs.~\cite{goltsev2008percolation} and \cite{goltsev2012localization}, since the nonbacktracking matrix is not symmetric, and so the full set of its eigenvectors does not form an orthogonal basis. To surpass this difficulty, we introduce a new symmetric matrix 
 \begin{equation}
 \mathbf{M} = \mathbf{A} - \lambda_1^{-1} (\mathbf{D} - \mathbf{I}) 
 , 
\label{eq:144}
\end{equation}
where $\lambda_1$ is the leading eigenvalue of the nonbacktracking matrix and matrices $\mathbf{A}$, $\mathbf{D}$, and $\mathbf{I}$ are the adjacency matrix, the degree matrix ($D_{ij}=q_i\delta_{ij}$), and the identity matrix, respectively. Matrix $\mathbf{M}$ is a special case of the Bethe Hessian matrix \cite{saade2014spectral}. Its eigenvectors already constitute a complete orthogonal basis. The leading eigenvalue of the nonbacktracking matrix, $\lambda_1$, is also an eigenvalue of $\mathbf{M}$, and the corresponding eigenvector components can be expressed in terms of the components of the principal eigenvector of the nonbacktracking matrix. These useful properties of matrix $\mathbf{M}$ enable us to calculate the slope of the order parameter at $p_c = \lambda_1^{-1}$,
(see the Appendix for details of the derivation):
\begin{widetext}
\begin{equation}
\Omega \cong  \frac{  \lambda_1 \left( \mathop{\sum}\limits_i \mathop{\sum}\limits_{j \in \partial_i} v^{(\lambda_1)}_{i\leftarrow j} \right)  \mathop{\sum}\limits_i (q_i{-}1 - {\lambda_1}^2) \left( \mathop{\sum}\limits_{j \in \partial_i} v^{(\lambda_1)}_{i\leftarrow j} \right)^2 }{ N \mathop{\sum}\limits_{i} \left( \mathop{\sum}\limits_{j \in \partial_i} v^{(\lambda_1)}_{i\leftarrow j} \right) \mathop{\sum}\limits_{j \in \partial_i} \left(  \mathop{\sum}\limits_{k,k'>k \in \partial_i \backslash j}  v^{(\lambda_1)}_{i\leftarrow k} v^{(\lambda_1)}_{i\leftarrow k'} - \lambda_1 \mathop{\sum}\limits_{k,k'>k \in \partial_j \backslash i}   v^{(\lambda_1)}_{j\leftarrow k} v^{(\lambda_1)}_{j\leftarrow k'}\right)}  .
\label{eq:145}
\end{equation}
\end{widetext}
Figure~\ref{fig:curves} compares the slopes of the order parameter $S(p)$ calculated in the critical region solving Eq.~(\ref{eq:143}) for the 4-clique, an Erd\H{o}s--R\'enyi random graph and three real networks with the slopes found by using formula~(\ref{eq:145}). Note the perfect agreement at $p_c$. The curves in Fig.~\ref{fig:curves}(b) also reveal the sizes of the regions in which the relation $S \cong \Omega (p-p_c)$ is applicable for these networks. 

The slope of the order parameter may also be obtained using a different method, which does not rely on symmetric matrices. This alternative method uses the fact that for any diagonalizable matrix, the complex conjugate of any left eigenvector corresponding to one eigenvalue is orthogonal to all right eigenvectors corresponding to a different eigenvalue (and vice versa, the conjugate of any right eigenvector corresponding to one eigenvalue is orthogonal to all left eigenvectors corresponding to a different eigenvalue). This method yields a simpler expression for the slope,
(see the Appendix for details of the derivation):
\begin{equation}
\Omega=\frac{{\lambda_1}^2\left( \sum_{i\leftarrow j} v^{\lambda_1}_{i\leftarrow j} \right)\left( \sum_{i\leftarrow j}  v^{\lambda_1}_{j\leftarrow i} v^{\lambda_1}_{i\leftarrow j} \right)} 
 {N \sum_{i\leftarrow j}  v^{\lambda_1}_{j\leftarrow i}  \sum_{k,k'>k \in \mathcal{N}_j \backslash i}  v^{\lambda_1}_{j\leftarrow k} v^{\lambda_1}_{j\leftarrow k'}}   .
\label{eq:145b}
\end{equation}
%
%
Both approaches have their respective merit and they provide equivalent expressions for the slope of the order parameter, Eqs. (\ref{eq:145}) and (\ref{eq:145b}).
The first method introduces a new symmetric matrix, which may be useful for various problems on undirected graphs, where the underlying matrix---the adjacency matrix---is symmetric.
The second approach requires only that the relevant matrix be diagonalizable, therefore it may be used to treat a wide range of problems in nonsymmetric systems, e.g., percolation on directed networks. A detailed discussion of these issues will be presented elsewhere.

The above methods can be applied to the general class of problems defined by the set of message passing equations
\begin{equation}
 a_{i \leftarrow j} = F_{i \leftarrow j}(p, \{ a_{j \leftarrow k} : k \in \partial j \setminus i \}),
 \label{eq:146}
\end{equation}
with the condition that
%
\begin{equation}
 \frac{\partial F_{i \leftarrow j}}{\partial a_{j \leftarrow k}} \Bigr|_{p=p_c} = c,
 \label{eq:147}
\end{equation}
where $c$ is a constant, independent of $i \leftarrow j$ and $j \leftarrow k$. For any such model the solutions $a_{i \leftarrow j}$ of Eq.~(\ref{eq:146}) and the order parameter $S$ near $p_c$ can be calculated using our schemes (see the Appendix).

As another example of the application of our results, we consider an algorithm 
approximating optimal percolation, suggested in
Ref.~\cite{morone2015influence}. It was shown that an  
efficient way of destroying a network (removing the lowest number of nodes in order
to disconnect the giant component) is by sequentially removing nodes according to their \textit{collective influence} $CI_\ell(i)$. This characteristic of 
node $i$ is defined as
\begin{equation}
   CI_\ell(i) = (q_i-1) \sum_{j \in \partial \text{Ball}(i,\ell)} (q_j-1)
  \label{eq:150}
  ,
\end{equation}
where $q_i$ is the degree of node $i$ and $\partial \text{Ball}(i,\ell)$ is the set of 
the nodes that are at a distance $\ell$ from node $i$.
In this algorithm, $\ell$ must be large but smaller than the diameter of the graph under consideration. 
Let us consider the same problem on the nonbacktracking expansion of a finite graph.
We see that in this case the collective influences $CI_\ell(i)$---taking $\ell \rightarrow \infty$---are just the asymptotic relative sizes of the surfaces
of $\mathcal{T}(\ell, i)$, multiplied by $q_i - 1$. 
We have already expressed these relative sizes in terms of the components of the principal eigenvector 
of the nonbacktracking matrix of the original graph [see Eq.~(\ref{eq:110})].
If the finite graph under consideration is large and has no short loops, we can assume that using our expression is a good approximation
to the optimal way of disconnecting the graph.
An improvement on the collective influence method is suggested in Ref.~\cite{morone2016collective}. It was shown empirically that better attack performance can be achieved by sequentially removing nodes according to the following centrality measure (collective influence propagation)
\begin{equation}
   CI_p(i) =  \sum_{j \in \partial i} w_{i \leftarrow j}^{(\lambda_1)} v_{i \leftarrow j}^{(\lambda_1)} + w_{j \leftarrow i}^{(\lambda_1)} v_{j \leftarrow i}^{(\lambda_1)}
   ,
  \label{eq:160}
\end{equation}
where $w_{i \leftarrow j}^{(\lambda_1)}$ and $v_{i \leftarrow j}^{(\lambda_1)}$ are the components of the principal left and right eigenvectors of the nonbacktracking matrix, respectively.
It was also shown in \cite{morone2016collective} analytically that removing a node according to Eq. (\ref{eq:160}) results in the biggest decrease in $\lambda_1$, the leading eigenvalue of the nonbacktracking matrix, i.e., the mean branching of the corresponding NBE. Thus, this method produces ``stepwise optimal'' percolation.
From the result $w_{j \leftarrow i}^{(\lambda_1)} = v_{i \leftarrow j}^{(\lambda_1)}$ we see that using the centrality index of Eq. (\ref{eq:160}) is actually equivalent to using
\begin{equation}
   CI_p(i) =  \sum_{j \in \partial i} w_{i \leftarrow j}^{(\lambda_1)} v_{i \leftarrow j}^{(\lambda_1)} = 
   \sum_{j \in \partial i} v_{i \leftarrow j}^{(\lambda_1)} v_{j \leftarrow i}^{(\lambda_1)}
   .
  \label{eq:162}
\end{equation}
Considering the nonbacktracking expansion we can give a physical interpretation of the product $w_{i \leftarrow j}^{(\lambda_1)} v_{i \leftarrow j}^{(\lambda_1)}$. This is the relative number of infinite paths in the NBE passing through a link $i \leftarrow j$ (or link $ j \leftarrow i$). The relative number of infinite paths arriving at a link $i \leftarrow j$ is given by $v_{i \leftarrow j}^{(\lambda_1)}$ and the relative number of infinite paths starting from a link $i \leftarrow j$ is given by $w_{i \leftarrow j}^{(\lambda_1)}$ (see Sec. \ref{s3}). Summing the product $w_{i \leftarrow j}^{(\lambda_1)} v_{i \leftarrow j}^{(\lambda_1)}$ over all neighbors $j$ of node $i$, we get the relative number of infinite paths in the NBE passing through a node $i$, which is exactly the $CI_p$ centrality index. 
One can consider the stepwise optimal removal of edges instead of nodes. For this problem, the expression (\ref{eq:162}) should be replaced by the edge centrality measure 
\begin{equation}
   CI_e(ij) = w_{i \leftarrow j}^{(\lambda_1)} v_{i \leftarrow j}^{(\lambda_1)} = v_{i \leftarrow j}^{(\lambda_1)} v_{j \leftarrow i}^{(\lambda_1)}
  \label{eq:165}
\end{equation}
for edge $ij$.
Note that in 
the recent work 
\cite{braunstein2016network} a method for finding the decycling number of graphs was presented,  outperforming all previous approaches to optimal percolation.
A computationally much more efficient method, providing a similarly high performance, was introduced in \cite{zdeborova2016fast}.


\section{Accuracy and limitations of the message passing approach}
\label{s5}

We have shown that when we are using message passing equations to approximately solve any given problem on a finite graph, we are actually solving the problem exactly on the nonbacktracking expansion of the given finite graph. At present there is no theory available to determine how good this approximation is in the general case. We now briefly discuss the accuracy of this approximation, through the example of random percolation, and how well it may be expected to work in certain situations.

Note that although we defined the NBE for finite graphs, it may also be defined for infinite networks as a limit of finite networks. From Refs. \cite{karrer2014percolation} and \cite{hamilton2014tight} we know that for infinite networks the critical point (for percolation) of the NBE is a lower bound on that of the original network. An even stronger statement is also true: the relative size of the giant component in the NBE is an upper bound on that of the original network (for any link activation probability $p$). This is easily seen considering the following. The solutions $H_{i \leftarrow j}(1)$ of Eq. (\ref{eq:140}) are a lower bound on the actual probabilities of reaching a finite cluster when following link $i \leftarrow j$ in the direction of $j$, due to the presence of loops in the original network (see \cite{karrer2014percolation}). According to Eq. (\ref{eq:143}), therefore, the order parameter of the NBE is an upper bound on that of the original network. With this result in mind we can quantify the \textit{badness} $R$ of the NBE approximation as
\begin{equation}
   R = \frac{\int_0^1 \lvert S_{NBE}(p) - S(p) \rvert dp}{\int_0^1 S(p) dp}
   ,
  \label{eq:170}
\end{equation}
%
%
%
where $S(p)$ is the order parameter of the original network and $S_{NBE}(p)$ is the order parameter of the corresponding NBE, for a given link activation probability $p$. For infinite networks the absolute value in Eq. (\ref{eq:170}) is not necessary, as $S_{NBE}(p) - S(p) > 0$ for any $p$ in this case. For finite graphs, however, $S_{NBE}(p) - S(p) < 0$ for $p < p_{c, NBE}$, where $p_{c, NBE}$ is the critical point of the NBE and $S(p)$ is the relative size of the largest component. Therefore the absolute value in Eq. (\ref{eq:170}) is required to make this a universally applicable, meaningful measure of the badness of the approximation.
For uncorrelated random networks the NBE coincides with the network itself (in the infinite size limit), therefore $R=0$.

\begin{figure}[t]
\centering
\includegraphics[width=7cm,angle=0.]{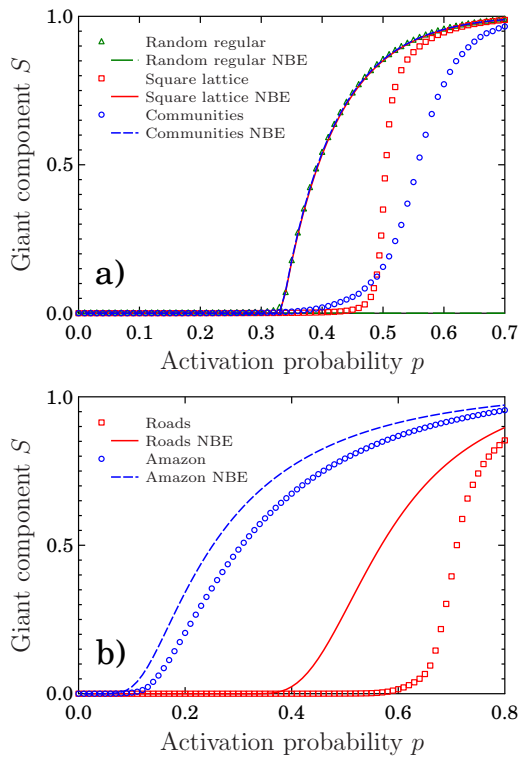}
\caption{(Color online) Simulation results and message passing solutions for the size of the giant connected component; examples of networks badly approximated by the nonbacktracking expansion. (a) Synthetic networks: a two-dimensional square lattice and a model network of maximal modularity $Q=1$. (b) Real-world networks: the road network of California \cite{network_roads} and an Amazon copurchase network \cite{network_amazon}. In all cases the simulation results differ significantly from the message passing solutions. See Table \ref{table1} for details.}
\label{fig:bad_fit}
\end{figure}

We now consider two representative examples of networks, having the same NBE, where the approximation fails: a two-dimensional square lattice and a synthetic modular network given by the following construction. Consider $n$ subnetworks, each of which is a random regular network of $m$ nodes and degree $4$. In each of the subnetworks, let us remove two randomly selected links, leaving four nodes of degree $3$. Then, let us connect these affected nodes to similar affected nodes in other subnetworks randomly, restoring the uniform degree $4$ for every node in the network. In the limit $n,m \to \infty$ this network has no finite loops and has modularity $Q=1$, according to the definition of modularity in Ref. \cite{newman2004finding}. The corresponding NBE is exactly the same as the one for a two-dimensional square lattice. This NBE also coincides with that of a $5$-clique; it is a random regular network of degree $4$. The above examples are particularly badly approximated by the NBE as shown in Fig. \ref{fig:bad_fit} (a).
%
%

%
\begin{table}
\centering
\begin{NoHyper}
\begin{tabular}{ |l|c|c|c| } 
 \hline
 \textbf{Network} & $N$ & $\langle q \rangle$ & $R$ \\
 \hline
 Random regular \footnotemark[1] & $10^5$ & $4$ & $0.00103$ \\
 \hline
 Square lattice \footnotemark[1] & $10^5$ & $4$ & $0.19752$ \\
 \hline
 Communities \footnotemark[1] & $10^5$ & $4$ & $0.31575$ \\
 \hline
 Roads \footnotemark[2] & $1965206$ & $2.82$ & $0.47778$ \\
 \hline
 Amazon\footnotemark[2]  & $334836$ & $5.53$ & $0.08517$ \\
 \hline
 C. elegans full \footnotemark[3] & $448$ & $21.17$ & $0.02765$ \\
 \hline
 C. elegans pharynx \footnotemark[3] & $51$ & $9.22$ & $0.01745$ \\
 \hline
 C. elegans main body \footnotemark[3] & $397$ & $22.69$ & $0.00380$ \\
 \hline
 Erd\H{o}s--R\'enyi \footnotemark[1] & $10^5$ & $5$ & $0.00064$ \\
 \hline
 Gnutella \footnotemark[2] & $62586$ & $4.73$ & $0.00037$ \\
 \hline
 Internet \footnotemark[4] & $22963$ & $4.22$ & $0.00071$ \\
 \hline
\end{tabular}
\footnotetext[1]{Computer generated.}
\footnotetext[2]{Downloaded from: \url{http://snap.stanford.edu/data/}.}
\footnotetext[3]{Downloaded from: \url{http://www.wormatlas.org/}.}
\footnotetext[4]{Downloaded from: \url{http://www-personal.umich.edu/~mejn/netdata/}.}

\end{NoHyper}

\caption{Size $N$ (number of nodes), mean degree $\langle q \rangle$, and badness value $R$ [defined in Eq. (\ref{eq:170})] for the networks considered in Sec. \ref{s5}.}
\label{table1}

\end{table}
These two examples represent two classes of networks that are not well approximated by the NBE: low-dimensional networks and highly modular networks.
Low-dimensional networks have many short loops and, therefore, are essentially non-treelike. A modification of the original message passing method has been suggested in \cite{radicchi2016beyond} to deal with clustering and in \cite{yoon2011belief} for networks with loopy motifs. Low dimensionality, however, implies many intertwined loops of many different sizes, which appears to be a significant factor in rendering such systems untreatable by the locally treelike approximation.
Interestingly, even if there are no finite loops in the network, the NBE may still be a bad approximation if the modularity is very high [see the model modular network ``Communities'' in Fig. \ref{fig:bad_fit} (a)].
A message passing method for modular networks is suggested in \cite{faqeeh2015emergence}, however, to use it, one needs to know the modular structure of the network \textit{a priori}. The problem is, that there is no unique definition of modularity in networks. Also, finding the community structure of a network---using any meaningful definition of modularity---typically leads to NP-hard decision problems \cite{schaeffer2007graph}.
In Fig. \ref{fig:bad_fit} (b) we present real-world examples of the two classes of networks discussed above: a planar network (road network of California \cite{network_roads}) and a modular network (Amazon copurchase network \cite{network_amazon}). The respective badness values are listed in Table \ref{table1}.

\begin{figure}[t]
\centering
\includegraphics[width=7cm,angle=0.]{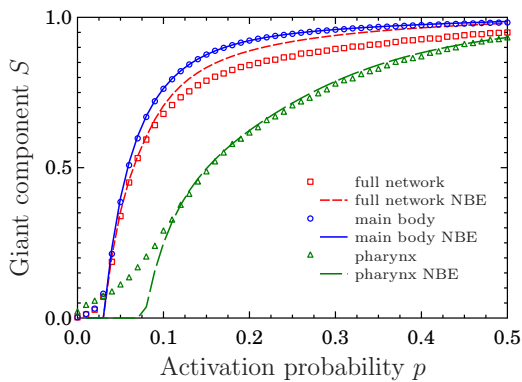}
\caption{(Color online) Simulation results and message passing solutions for the size of the giant connected component for the neural network of \textit{C. elegans} \cite{jarrell2012connectome}. The full neural network consists of two well-defined modules: the main body and the pharynx. The message passing equations are a bad approximation for the network as a whole, but a surprisingly good one for the modules separately. (Note the very small size, $N=51$, of the pharynx network!)}
\label{fig:C_elegans}
\end{figure}

An instructive example that simultaneously demonstrates the pitfalls and the remarkable power of this approximation is the neural network of the roundworm \textit{Caenorhabditis elegans} (\textit{C. elegans}) \cite{jarrell2012connectome}. We investigated the hermaphrodite \textit{C. elegans}, combining both chemical and electrical synapses and removing multiple links and self-loops. This neural network contains two obvious communities: the pharynx and the main body of the roundworm. The two communities are separated by only two links, hence the obvious modular structure of the network. Figure \ref{fig:C_elegans} shows percolation simulation results and the corresponding message passing solutions for the whole network and also for the two subnetworks (obtained by cutting the two links separating the modules). The NBE approximation is not particularly good when the whole network is considered, but is remarkably accurate for the two subnetworks. These subnetworks are very small, and have high clustering coefficients ($C = 0.47$ for the pharynx and $C = 0.26$ for the main body), implying that modularity is a much more important factor in determining the accuracy of the approximation.

\begin{figure}[t]
\centering
\includegraphics[width=7cm,angle=0.]{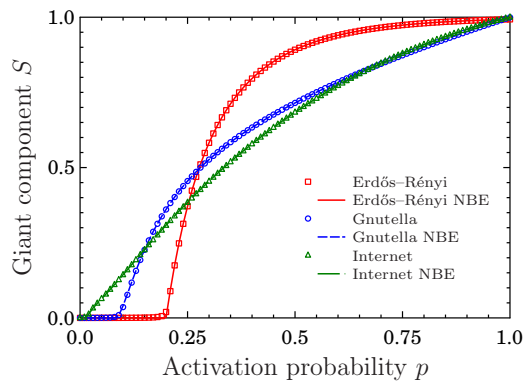}
\caption{(Color online) Simulation results and message passing solutions for networks where the nonbacktracking expansion gives an excellent approximation: an Erd\H{o}s--R\'enyi network, a subgraph of the Gnutella peer-to-peer file sharing network \cite{network_gnutella}, and a snapshot of the Internet at the level of autonomous systems \cite{meyer2001university}. The simulation results practically coincide with the message passing curves.}
\label{fig:perfect_fit}
\end{figure}

Figure \ref{fig:perfect_fit} shows ``less correlated'' networks---also studied in \cite{karrer2014percolation}---where the badness values of the approximation are very close to $0$ (see Table \ref{table1}). These networks are small worlds, i.e., infinite-dimensional, and have low modularities. We showed above that low dimensionality and high modularity are two attributes that cause the approximation to fail. Modular networks are ubiquitous in nature, therefore such excellent agreement as in Fig. \ref{fig:perfect_fit} (or Ref. \cite{karrer2014percolation}) may not actually be a very common case.
%

Finally, we discuss another limitation of message passing methods, their time complexity. A significant advantage of using message passing equations as opposed to doing simulations is that, supposedly, the computational time required to find the solutions is much less than the time involved in doing the actual simulations and averaging over many realizations. This is not always true, however. The number of operations required in message passing methods (i.e., when solving the message passing equations by iterations) is
\begin{equation}
   T_{MP} \sim n_{conv} N \langle q \rangle \sim n_{conv} L
   ,
  \label{eq:175}
\end{equation}
where the number of iterations to convergence is denoted $n_{conv}$, and $L$ is the number of links in the network.
To achieve this complexity one may do the following. Considering the percolation problem, Eq. \ref{eq:140}, at each iteration, for every node, first compute the product of all incoming messages. This operation has time complexity $\sim L$. To update each outgoing message from every node, divide the precalculated product for the starting node of this message by the incoming message for the same link. This, again, has complexity $\sim L$, resulting in the overall complexity given in Eq. (\ref{eq:175}).
In order for the message passing equations to converge to a solution, the messages at any part of the network must have information from every other part of the network. Therefore, $n_{conv}$ cannot be less than the network diameter, i.e., the longest of all the shortest paths between node pairs in the network. In a typical network demonstrating the small-world effect the diameter grows logarithmically with the system size $N$. In this case $n_{conv}$ can be expected to grow at least as $n_{conv} \sim \ln N$, and so for networks with sufficiently rapidly decaying degree distributions $T_{MP} \sim N \ln N$.
(Note that in networks with a divergent second moment of the degree distribution, the diameter may increase with $N$ even more slowly than logarithmically \cite{cohen2003scale, dorogovtsev2003metric}. At the other extreme, namely in the case of a chain, the diameter grows as $N$.)
%
%
Let us compare these results with the time complexity of doing actual simulations.

Say we want to determine the size of the ``giant'' percolating cluster in a large graph to a given accuracy (given standard error). The quantity we are interested in is an average over $n$ samples, $S = \langle s \rangle$, where $s$ is a binary quantity ($s=1$ or $s=0$) indicating whether or not a given node of the graph is in the giant component. The standard error of the estimate for $S$ is simply $\sigma / \sqrt{n}$, where $\sigma$ is the standard deviation of $s$. Therefore, for a given desired error in $S$ the number of node samples needed is independent of the system size. To determine whether a node belongs to the giant component, we must explore the neighborhood of this node (e.g., using a simple breadth first search) visiting up to $l$ other nodes, where $l \gg l^*$. Here $l^*$ is the typical size of a finite cluster. The typical cluster size depends on the distance from the critical point but not the system size. Considering the above, we arrive at the following conclusion. At a given distance from the critical point, for a given desired accuracy, the order parameter can be determined using simulations in constant time for any network architecture. This is in strong contrast with the message passing approach, where the accuracy is unknown and the time complexity significantly higher.
While the message passing method may be faster in certain small networks, it is always inferior to simple simulations for large enough networks.


\section{Discussion and extensions}
\label{s6}

Message passing equations often provide a good approximation to the behavior of interacting models on large sparse networks. In this paper, for any finite graph, we have presented a network construction for which the message passing equations are exact, thereby giving a physical interpretation of this approximation. For a wide range of problems, represented by message passing equations, we have shown how to express the solutions near the critical point in terms of the principal eigenvector components of the nonbacktracking matrix.

It is well established that this approximation should work well for infinite-dimensional (small-world) networks which are sparse and have no strong degree-degree correlations. It is not clear how well it should work for networks of low dimensionality, many loops, and correlations. We have introduced a way of quantifying the accuracy of the approximation and elaborated on two distinct classes of networks where this approach fails: low-dimensional and highly modular networks. As real-world examples of these cases we considered the road network of California and an Amazon copurchase network, where the message passing equations, indeed, proved a bad approximation.

Message passing methods based on the locally treelike approximation have a wide range of applicability and, in some cases, provide good results. However, as we have shown, for increasingly large systems these methods have a high computational cost compared with simple simulations. Also, while one can always find an estimate of the error associated with simulations in a straightforward way, the accuracy of message passing methods is entirely unknown.

All the results in this paper apply also to weighted graphs. In this case one must work with a weighted nonbacktracking matrix, obtained by multiplying each row of the unweighted nonbacktracking matrix by the weight of the corresponding link. Using this modified matrix, the critical point and the solutions near the critical point are determined exactly as in the unweighted case.

The network construction presented here may be generalized straightforwardly to directed graphs. The directed percolation critical point is also given by the inverse of the leading eigenvalue of the nonbacktracking matrix. Solutions near the critical point can be found using the method we introduced for nonsymmetric systems (see Method 2 in the Appendix).

The nonbacktracking expansion can be generalized to various percolation problems for interdependent and multiplex networks \cite{buldyrev2010catastrophic, bianconi2014multiple, boccaletti2014structure, kivela2014multilayer, radicchi2015percolation, radicchi2017redundant, baxter2016correlated}.
In the case of multiplex networks, the only difference from single-layer (simplex) networks is that the resulting infinite network (the NBE) will consist of links of different colors, corresponding to links of different layers. The message passing equations written for any given model on finite multiplex networks \cite{bianconi2014multiple} will be exact in the corresponding NBE, just as in simplex networks.
A number of percolation problems on multiplex and interdependent networks exhibit discontinuous transitions. In order to retain the interesting physics in these systems, one has to account for the nonlinear terms in the basic equations \cite{radicchi2015percolation}. Finding the analytical solutions similar to our Eqs. (\ref{eq:145}) and (\ref{eq:145b}) at discontinuous transitions in multiplex networks is an interesting challenge for the future.

\vskip 0.3cm

We thank L. Zdeborov{\'a} for useful comments.
This work was supported by the FET proactive IP project MULTIPLEX 317532.


\begin{widetext}

\section*{Appendix: Finding solutions near $p_c$}
\label{app:main}
\setcounter{equation}{0}
\renewcommand{\theequation}{A\arabic{equation}}

We stated that in the class of physical problems on undirected networks defined by Eqs.~(\ref{eq:146}) and (\ref{eq:147}) the solution near $p_c$ can be expressed in terms of the components of the principal eigenvector of the nonbacktracking matrix. Here we derive this expression in the particular case of the message passing equations, (\ref{eq:140}), for percolation.  

We expand the message passing equations, Eq. (\ref{eq:140}), as
\begin{equation}
\!\!
a_{i\leftarrow j} = p\!\left(\sum_{k\in \partial j \backslash i} \!\!\! a_{j\leftarrow k} -\!\!\!\!\! \sum_{k,k'>k \in \partial j \backslash i} \!\!\!\!\!\!  a_{j\leftarrow k} a_{j\leftarrow k'} + \dots \! \right)\!
,
\label{eq1}
\end{equation}
where $a_{i\leftarrow j}=1-H_{i\leftarrow j}(1)$. 
The set of eigenvectors of the nonbacktracking matrix, $\mathbf{v}^{(\lambda)}$, forms a complete basis enabling us to express the vector $\mathbf{a}=\{a_{i\leftarrow j}\}$ as a linear combination of the eigenvectors 
\begin{equation}
\mathbf{a}=\sum_\lambda C_\lambda \mathbf{v}^{(\lambda)}
.
\label{eq2}
\end{equation}
%
Notice that both $\mathbf{a}$ and the coefficients $C_\lambda$ vary with $p$, approaching $0$ when $\delta=p-p_c \to 0$.
Replacing $a_{i\leftarrow j}$ with $\sum_\lambda C_\lambda v^{(\lambda)}_{i\leftarrow j}$ in Eq.~(\ref{eq1}), we obtain the equation 
%
\begin{equation}
\sum_\lambda C_\lambda v^{(\lambda)}_{i\leftarrow j} (p\lambda-1) = p \sum_{\lambda, \lambda'} C_\lambda C_{\lambda'} \!\!\!\!\!\!\!\! \sum_{k,k'>k \in \partial j \backslash i} \!\!\!\!\!\! v^{(\lambda)}_{j\leftarrow k} v^{(\lambda')}_{j\leftarrow k'}{+}\dots ,
\label{eq3}
\end{equation}
%
where we have used the identity 
$$
\sum_{k\in \partial _j \backslash i} v^{(\lambda)}_{j\leftarrow k}=\lambda v^{(\lambda)}_{i\leftarrow j},
$$  
since $\mathbf{v}^{(\lambda)}$ is the eigenvector of the nonbacktracking matrix with eigenvalue $\lambda$. 
Since all $C_\lambda$ approach $0$ as $\delta \to 0$, Eq.~(\ref{eq3}) implies that there is an eigenvalue $\lambda_1=1/p_c$, such that $C_{\lambda_1}= O(\delta)$, and that for $\lambda\neq\lambda_1$ the coefficients $C_{\lambda}$ approach $0$ as $\delta^2$ or more rapidly.
Furthermore, by the Perron-Frobenius theorem if all components of $\mathbf{v}^{(\lambda_1)}$ are non-negative, then the eigenvalue $\lambda_1$ is positive real and has the largest absolute value (hence, the subscript $1$). 
In this way we find that at small $\delta$

\begin{equation}
a_{i\leftarrow j} \cong C_{\lambda_1} v^{(\lambda_1)}_{i\leftarrow j},
\label{eq4}
\end{equation}
where 
$v^{(\lambda_1)}_{i\leftarrow j}$ are the components of the principal eigenvector of the nonbacktracking matrix.
From this point onwards one may use two different approaches. \textit{Method 1} introduces an auxiliary symmetric matrix, to proceed with the derivation in the standard way. \textit{Method 2} exploits a useful property of the left and right eigenvectors of diagonalizable matrices, resulting in a much shorter derivation and a more compact final result.

\subsection*{Method 1}

Let us consider the following sums, over the set of nearest neighbors $\partial i$ of node $i$,
\begin{equation}
b_{i\leftarrow} = \sum_{j \in \partial i} a_{i\leftarrow j} \text{\ \ \ \ and \ \ \ \ } b_{\leftarrow i} = \sum_{j \in \partial i} a_{j\leftarrow i}.
\label{eq5}
\end{equation}
We sum both sides of Eq.~(\ref{eq1}):
\begin{eqnarray}
&&b_{i\leftarrow} {=} p\!\left(\sum_{j \in \partial_i}  \! b_{j\leftarrow} {-} b_{\leftarrow i} {-}\!\! \sum_{j \in \partial_i}\ \sum_{k,k'>k \in \mathcal{N}_j \backslash i} \!\!\!\!\!\!\!  a_{j\leftarrow k} a_{j\leftarrow k'} {+} \dots\! \right)\!\!,
\nonumber
\\
&&b_{\leftarrow i} {=} p\!\left( (q_i{-}1) b_{i\leftarrow} {-}\!\! \sum_{j \in \partial_i}\ \sum_{k,k'>k \in \partial_i \backslash j} \!\!\!\!\!\!\!  a_{i\leftarrow k} a_{i\leftarrow k'} {+} \dots\! \right)\!\!.
\label{eq6}
\end{eqnarray}
Replacing $b_{\leftarrow i}$ in the first of these equations with the right-hand side of the second equation we obtain
%
\begin{equation}
b_{i\leftarrow} = p\left[\sum_{j \in \partial_i}   b_{j\leftarrow} -  p (q_i{-}1) b_{i\leftarrow}  + \sum_{j \in \partial_i} \left( p \sum_{k,k'>k \in \partial_i \backslash j}  a_{i\leftarrow k} a_{i\leftarrow k'} -  \sum_{k,k'>k \in \partial_j \backslash i}   a_{j\leftarrow k} a_{j\leftarrow k'}\right) + \dots \right],
\label{eq7}
\end{equation}
%
where $q_i$ is the degree of node $i$.
Let us introduce the matrix $\mathbf{M}$ with elements $M_{ij}$ defined as 
\begin{equation}
M_{ij}=
\begin{cases}
A_{ij} &\text{if } i\neq j, 
\\
-p_c(q_i-1) &\text{if } i=j
,
\end{cases}
\label{eq8}
\end{equation}
where $A_{ij}$ are the elements of the adjacency matrix.
The matrix $\mathbf{M}$ is symmetric, and so we can express $\mathbf{b}=\{b_{1\leftarrow}, b_{2\leftarrow}, \ldots, b_{N\leftarrow}\}$ as a linear combination of its eigenvectors, $\mathbf{u}^{(\tilde{\lambda})}$, which form a complete orthogonal basis, 
\begin{equation}
\mathbf{b} = \sum_{\tilde{\lambda}} C_{\tilde{\lambda}} \mathbf{u}^{(\tilde{\lambda})}.
\label{eq9}
\end{equation}
In the following we denote the eigenvalues of $\mathbf{M}$ by $\tilde{\lambda}$. These eigenvalues are different from the eigenvalues $\lambda$ of the nonbacktracking matrix. We replace $b_{i\leftarrow}$ and $a_{i\leftarrow j}$ in Eq.~(\ref{eq7}) with their decompositions, Eqs.~(\ref{eq9}) and (\ref{eq4}), respectively,
%
\begin{equation}
\sum_{\tilde{\lambda}} C_{\tilde{\lambda}} u^{(\tilde{\lambda})}_{i} (p_c\tilde{\lambda}{-}1) {-} \delta  \sum_{\tilde{\lambda}} [p_c(q_i{-}1){-}\tilde{\lambda}] C_{\tilde{\lambda}} u^{(\tilde{\lambda})}_{i}  {=}  {-} p_c\sum_{\lambda , \lambda' } C_{\lambda} C_{\lambda'} \sum_{j \in \partial_i}\!\! \left( p_c \!\!\!\!\!\! \sum_{k,k'>k \in \partial_i \backslash j} \!\!\!\!\!\!\!\! v^{(\lambda)}_{i\leftarrow k} v^{(\lambda')}_{i\leftarrow k'} {-} \!\!\!\!\!\!\! \sum_{k,k'>k \in \partial_j \backslash i} \!\!\!\!\!\!\!\!  v^{(\lambda)}_{j\leftarrow k} v^{(\lambda')}_{j\leftarrow k'}\right){+} \dots .
\label{eq10}
\end{equation}
%
This equation implies that among the eigenvalues $\tilde{\lambda}$ there is one, say $\tilde{\lambda}^*$, equal to $1/p_c=\lambda_1$. Note that the eigenvalue $\tilde{\lambda}^*$ is not the leading eigenvalue of $\mathbf{M}$. 
Note also that according to Eqs.~(\ref{eq4}), (\ref{eq5}), and (\ref{eq9}), $C_{\tilde{\lambda}^*} u^{(\tilde{\lambda}^*)}_{i} = C_{\lambda_1}  \sum_{j \in \partial_i} v^{(\lambda_1)}_{i\leftarrow j}\sim \delta$, while the rest of the coefficients $C_\lambda$ and $C_{\tilde{\lambda}}$ approach $0$ as $\delta^2$ or more rapidly.
To obtain $C_{\lambda_1}$ we multiply both sides of Eq.~(\ref{eq10}) by $u^{(\tilde{\lambda}^*)}_{i}$ and sum over $i$. The result is 
%
\begin{equation}
 {C_{\lambda_1}} \cong  \frac{   \lambda_1 \mathop{\sum}\limits_i (q_i{-}1 - {\lambda_1}^2)  \left( \mathop{\sum}\limits_{j \in \partial_i} v^{(\lambda_1)}_{i\leftarrow j} \right)^2}{ \mathop{\sum}\limits_{i} \left( \mathop{\sum}\limits_{j \in \partial_i} v^{(\lambda_1)}_{i\leftarrow j} \right) \mathop{\sum}\limits_{j \in \partial_i} \left(  \mathop{\sum}\limits_{k,k'>k \in \partial_i \backslash j}  v^{(\lambda_1)}_{i\leftarrow k} v^{(\lambda_1)}_{i\leftarrow k'} - \lambda_1 \mathop{\sum}\limits_{k,k'>k \in \partial_j \backslash i}   v^{(\lambda_1)}_{j\leftarrow k} v^{(\lambda_1)}_{j\leftarrow k'}\right)}  \delta .
\label{eq11}
\end{equation}
%
Note that the contribution of the first term on the left-hand side of Eq.~(\ref{eq10}) is $0$ because the eigenvectors $\mathbf{u}^{(\tilde{\lambda})}$ are orthogonal. 
Finally, we express the order parameter near $p_c$ in terms of the principal eigenvector components of the nonbacktracking matrix as
\begin{eqnarray}
S &=&\frac{1}{N}\sum_i \left(1-\prod_{j \in \partial_i} H_{i \leftarrow j}(1)\right)\cong \frac{1}{N}\sum_i \sum_{j \in \partial_i} a_{i \leftarrow j}
\nonumber
\\
&\cong& \frac{C_{\lambda_1}}{N}\sum_i \sum_{j \in \partial_i} v^{(\lambda_1)}_{i \leftarrow j} = \Omega \cdot (p-p_c),
\label{eq12}
\end{eqnarray}
where 
%
\begin{equation}
\Omega \cong  \frac{  \lambda_1 \left( \mathop{\sum}\limits_i \mathop{\sum}\limits_{j \in \partial_i} v^{(\lambda_1)}_{i\leftarrow j} \right)  \mathop{\sum}\limits_i (q_i{-}1 - {\lambda_1}^2) \left( \mathop{\sum}\limits_{j \in \partial_i} v^{(\lambda_1)}_{i\leftarrow j} \right)^2 }{ N \mathop{\sum}\limits_{i} \left( \mathop{\sum}\limits_{j \in \partial_i} v^{(\lambda_1)}_{i\leftarrow j} \right) \mathop{\sum}\limits_{j \in \partial_i} \left(  \mathop{\sum}\limits_{k,k'>k \in \partial_i \backslash j}  v^{(\lambda_1)}_{i\leftarrow k} v^{(\lambda_1)}_{i\leftarrow k'} - \lambda_1 \mathop{\sum}\limits_{k,k'>k \in \partial_j \backslash i}   v^{(\lambda_1)}_{j\leftarrow k} v^{(\lambda_1)}_{j\leftarrow k'}\right)}  .
\label{eq13}
\end{equation}
%

\subsection*{Method 2}

Let us denote the right eigenvectors of the nonbacktracking matrix $\mathbf{v}^{(\lambda)}$ and the left eigenvectors $\mathbf{w}^{(\lambda)}$.
The complex conjugate of any $\mathbf{v}^{(\lambda)}$ is orthogonal to all $\mathbf{w}^{(\lambda')}$ ($\lambda \neq \lambda'$), and the conjugate of any $\mathbf{w}^{(\lambda)}$ is orthogonal to all $\mathbf{v}^{(\lambda')}$ ($\lambda \neq \lambda'$). [Indeed, let $\lambda \neq \lambda'$ be two different eigenvalues of $\mathbf{B}$. Then clearly $\lambda \mathbf{w}^{(\lambda)} \mathbf{v}^{(\lambda')} = \mathbf{w}^{(\lambda)} \mathbf{B} \mathbf{v}^{(\lambda')} = \lambda' \mathbf{w}^{(\lambda)} \mathbf{v}^{(\lambda')}$, giving $(\lambda - \lambda') \mathbf{w}^{(\lambda)} \mathbf{v}^{(\lambda')} = 0$, and therefore, $\mathbf{w}^{(\lambda)} \mathbf{v}^{(\lambda')} = 0$.]

Recall Eq. (\ref{eq3}),
\begin{equation}
\sum_\lambda C_\lambda v^{(\lambda)}_{i\leftarrow j} (p\lambda-1) = p \sum_{\lambda, \lambda'} C_\lambda C_{\lambda'} \!\!\!\!\!\!\!\! \sum_{k,k'>k \in \partial j \backslash i} \!\!\!\!\!\! v^{(\lambda)}_{j\leftarrow k} v^{(\lambda')}_{j\leftarrow k'}{+}\dots ,
\nonumber
\end{equation}
and that $v^{(\lambda)}_{i\leftarrow j}$ are components of the right eigenvector of the nonbacktracking matrix, corresponding to eigenvalue $\lambda$.
Multiplying both sides by $w^{\lambda_1}_{i\leftarrow j} $ and summing over $i\leftarrow j$ we see that the terms of $\lambda\neq\lambda_1$ become $0$. This procedure results in a considerably simpler expression for the coefficient $C_{\lambda_1}$,
\begin{equation}
 C_{\lambda_1} \cong \frac{{\lambda_1}^2 \sum_{i\leftarrow j}  v^{(\lambda_1)}_{j\leftarrow i} v^{(\lambda_1)}_{i\leftarrow j}} 
 { \sum_{i\leftarrow j}  v^{(\lambda_1)}_{j\leftarrow i}  \sum_{k,k'>k \in \mathcal{N}_j \backslash i}  v^{(\lambda_1)}_{j\leftarrow k} v^{(\lambda_1)}_{j\leftarrow k'}}  \delta ,
\label{eq15}
\end{equation}
where we have used the property of the nonbacktracking matrix of undirected graphs $w^{(\lambda)}_{i\leftarrow j}=v^{(\lambda)}_{j\leftarrow i}$, for simplification.
The singularity of the giant component $S=\Omega\delta$ has an amplitude, Eq.~(\ref{eq12}):
\begin{equation}
\Omega=\frac{{\lambda_1}^2\left( \sum_{i\leftarrow j} v^{(\lambda_1)}_{i\leftarrow j} \right)\left( \sum_{i\leftarrow j}  v^{(\lambda_1)}_{j\leftarrow i} v^{(\lambda_1)}_{i\leftarrow j} \right)} 
 {N \sum_{i\leftarrow j}  v^{(\lambda_1)}_{j\leftarrow i}  \sum_{k,k'>k \in \mathcal{N}_j \backslash i}  v^{(\lambda_1)}_{j\leftarrow k} v^{(\lambda_1)}_{j\leftarrow k'}}   .
\label{eq16}
\end{equation}
Equations~(\ref{eq15}) and~(\ref{eq16}) are equivalent to the expressions, Eqs.~(\ref{eq11}) and~(\ref{eq13}).

\end{widetext}


\begin{thebibliography}{40}%
\makeatletter
\providecommand \@ifxundefined [1]{%
 \@ifx{#1\undefined}
}%
\providecommand \@ifnum [1]{%
 \ifnum #1\expandafter \@firstoftwo
 \else \expandafter \@secondoftwo
 \fi
}%
\providecommand \@ifx [1]{%
 \ifx #1\expandafter \@firstoftwo
 \else \expandafter \@secondoftwo
 \fi
}%
\providecommand \natexlab [1]{#1}%
\providecommand \enquote  [1]{``#1''}%
\providecommand \bibnamefont  [1]{#1}%
\providecommand \bibfnamefont [1]{#1}%
\providecommand \citenamefont [1]{#1}%
\providecommand \href@noop [0]{\@secondoftwo}%
\providecommand \href [0]{\begingroup \@sanitize@url \@href}%
\providecommand \@href[1]{\@@startlink{#1}\@@href}%
\providecommand \@@href[1]{\endgroup#1\@@endlink}%
\providecommand \@sanitize@url [0]{\catcode `\\12\catcode `\$12\catcode
  `\&12\catcode `\#12\catcode `\^12\catcode `\_12\catcode `\%12\relax}%
\providecommand \@@startlink[1]{}%
\providecommand \@@endlink[0]{}%
\providecommand \url  [0]{\begingroup\@sanitize@url \@url }%
\providecommand \@url [1]{\endgroup\@href {#1}{\urlprefix }}%
\providecommand \urlprefix  [0]{URL }%
\providecommand \Eprint [0]{\href }%
\providecommand \doibase [0]{http://dx.doi.org/}%
\providecommand \selectlanguage [0]{\@gobble}%
\providecommand \bibinfo  [0]{\@secondoftwo}%
\providecommand \bibfield  [0]{\@secondoftwo}%
\providecommand \translation [1]{[#1]}%
\providecommand \BibitemOpen [0]{}%
\providecommand \bibitemStop [0]{}%
\providecommand \bibitemNoStop [0]{.\EOS\space}%
\providecommand \EOS [0]{\spacefactor3000\relax}%
\providecommand \BibitemShut  [1]{\csname bibitem#1\endcsname}%
\let\auto@bib@innerbib\@empty
\bibitem [{\citenamefont {Baxter}(2007)}]{baxter2007exactly}%
  \BibitemOpen
  \bibfield  {author} {\bibinfo {author} {\bibfnamefont {R.~J.}\ \bibnamefont
  {Baxter}},\ }\href@noop {} {\emph {\bibinfo {title} {Exactly Solved Models in
  Statistical Mechanics}}}\ (\bibinfo  {publisher} {Courier Corp., Mineola,
  NY},\ \bibinfo {year} {2007})\BibitemShut {NoStop}%
\bibitem [{\citenamefont {Bender}\ and\ \citenamefont
  {Canfield}(1978)}]{bender1978asymptotic}%
  \BibitemOpen
  \bibfield  {author} {\bibinfo {author} {\bibfnamefont {E.~A.}\ \bibnamefont
  {Bender}}\ and\ \bibinfo {author} {\bibfnamefont {E.~R.}\ \bibnamefont
  {Canfield}},\ }\bibfield  {title} {\enquote {\bibinfo {title} {The asymptotic
  number of labeled graphs with given degree sequences},}\ }\href@noop {}
  {\bibfield  {journal} {\bibinfo  {journal} {J. Combin. Theory Ser. A}\
  }\textbf {\bibinfo {volume} {24}},\ \bibinfo {pages} {296} (\bibinfo {year}
  {1978})}\BibitemShut {NoStop}%
\bibitem [{\citenamefont {Bollob{\'a}s}(1980)}]{bollobas1980probabilistic}%
  \BibitemOpen
  \bibfield  {author} {\bibinfo {author} {\bibfnamefont {B.}~\bibnamefont
  {Bollob{\'a}s}},\ }\bibfield  {title} {\enquote {\bibinfo {title} {A
  probabilistic proof of an asymptotic formula for the number of labelled
  regular graphs},}\ }\href@noop {} {\bibfield  {journal} {\bibinfo  {journal}
  {Eur. J. Combin.}\ }\textbf {\bibinfo {volume} {1}},\ \bibinfo {pages} {311}
  (\bibinfo {year} {1980})}\BibitemShut {NoStop}%
\bibitem [{\citenamefont {Newman}\ \emph {et~al.}(2001)\citenamefont {Newman},
  \citenamefont {Strogatz},\ and\ \citenamefont {Watts}}]{newman2001random}%
  \BibitemOpen
  \bibfield  {author} {\bibinfo {author} {\bibfnamefont {M.~E.~J.}\
  \bibnamefont {Newman}}, \bibinfo {author} {\bibfnamefont {S.~H.}\
  \bibnamefont {Strogatz}}, \ and\ \bibinfo {author} {\bibfnamefont {D.~J.}\
  \bibnamefont {Watts}},\ }\bibfield  {title} {\enquote {\bibinfo {title}
  {Random graphs with arbitrary degree distributions and their applications},}\
  }\href@noop {} {\bibfield  {journal} {\bibinfo  {journal} {Phys. Rev. E}\
  }\textbf {\bibinfo {volume} {64}},\ \bibinfo {pages} {026118} (\bibinfo
  {year} {2001})}\BibitemShut {NoStop}%
\bibitem [{\citenamefont {Karrer}\ \emph {et~al.}(2014)\citenamefont {Karrer},
  \citenamefont {Newman},\ and\ \citenamefont
  {Zdeborov{\'a}}}]{karrer2014percolation}%
  \BibitemOpen
  \bibfield  {author} {\bibinfo {author} {\bibfnamefont {B.}~\bibnamefont
  {Karrer}}, \bibinfo {author} {\bibfnamefont {M.~E.~J.}\ \bibnamefont
  {Newman}}, \ and\ \bibinfo {author} {\bibfnamefont {L.}~\bibnamefont
  {Zdeborov{\'a}}},\ }\bibfield  {title} {\enquote {\bibinfo {title}
  {Percolation on sparse networks},}\ }\href@noop {} {\bibfield  {journal}
  {\bibinfo  {journal} {Phys. Rev. Lett.}\ }\textbf {\bibinfo {volume} {113}},\
  \bibinfo {pages} {208702} (\bibinfo {year} {2014})}\BibitemShut {NoStop}%
\bibitem [{\citenamefont {Morone}\ and\ \citenamefont
  {Makse}(2015)}]{morone2015influence}%
  \BibitemOpen
  \bibfield  {author} {\bibinfo {author} {\bibfnamefont {F.}~\bibnamefont
  {Morone}}\ and\ \bibinfo {author} {\bibfnamefont {H.~A.}\ \bibnamefont
  {Makse}},\ }\bibfield  {title} {\enquote {\bibinfo {title} {Influence
  maximization in complex networks through optimal percolation},}\ }\href@noop
  {} {\bibfield  {journal} {\bibinfo  {journal} {Nature}\ }\textbf {\bibinfo
  {volume} {524}},\ \bibinfo {pages} {65} (\bibinfo {year} {2015})}\BibitemShut
  {NoStop}%
\bibitem [{\citenamefont {Morone}\ \emph {et~al.}(2016)\citenamefont {Morone},
  \citenamefont {Min}, \citenamefont {Bo}, \citenamefont {Mari},\ and\
  \citenamefont {Makse}}]{morone2016collective}%
  \BibitemOpen
  \bibfield  {author} {\bibinfo {author} {\bibfnamefont {F.}~\bibnamefont
  {Morone}}, \bibinfo {author} {\bibfnamefont {B.}~\bibnamefont {Min}},
  \bibinfo {author} {\bibfnamefont {L.}~\bibnamefont {Bo}}, \bibinfo {author}
  {\bibfnamefont {R.}~\bibnamefont {Mari}}, \ and\ \bibinfo {author}
  {\bibfnamefont {H.~A.}\ \bibnamefont {Makse}},\ }\bibfield  {title} {\enquote
  {\bibinfo {title} {Collective influence algorithm to find influencers via
  optimal percolation in massively large social media},}\ }\href@noop {}
  {\bibfield  {journal} {\bibinfo  {journal} {Sci. Rep.}\ }\textbf {\bibinfo
  {volume} {6}},\ \bibinfo {pages} {30062} (\bibinfo {year}
  {2016})}\BibitemShut {NoStop}%
\bibitem [{\citenamefont {Bau}\ \emph {et~al.}(2002)\citenamefont {Bau},
  \citenamefont {Wormald},\ and\ \citenamefont {Zhou}}]{bau2002decycling}%
  \BibitemOpen
  \bibfield  {author} {\bibinfo {author} {\bibfnamefont {S.}~\bibnamefont
  {Bau}}, \bibinfo {author} {\bibfnamefont {N.~C.}\ \bibnamefont {Wormald}}, \
  and\ \bibinfo {author} {\bibfnamefont {S.}~\bibnamefont {Zhou}},\ }\bibfield
  {title} {\enquote {\bibinfo {title} {Decycling numbers of random regular
  graphs},}\ }\href@noop {} {\bibfield  {journal} {\bibinfo  {journal} {Random
  Struct. Algor.}\ }\textbf {\bibinfo {volume} {21}},\ \bibinfo {pages} {397}
  (\bibinfo {year} {2002})}\BibitemShut {NoStop}%
\bibitem [{\citenamefont {Hashimoto}(1989)}]{hashimoto2014automorphic}%
  \BibitemOpen
  \bibfield  {author} {\bibinfo {author} {\bibfnamefont {K.}~\bibnamefont
  {Hashimoto}},\ }\bibfield  {title} {\enquote {\bibinfo {title} {Zeta
  functions of finite graphs and representations of p-adic groups},}\ }in\
  \href@noop {} {\emph {\bibinfo {booktitle} {Automorphic Forms and Geometry of
  Arithmetic Varieties: Advanced Studies in Pure Mathematics}}},\ Vol.~\bibinfo
  {volume} {15},\ \bibinfo {editor} {edited by\ \bibinfo {editor}
  {\bibfnamefont {K.}~\bibnamefont {Hashimoto}}\ and\ \bibinfo {editor}
  {\bibfnamefont {Y.}~\bibnamefont {Namikawa}}}\ (\bibinfo  {publisher}
  {Kinokuniya},\ \bibinfo {address} {Tokyo},\ \bibinfo {year} {1989})\ p.\
  \bibinfo {pages} {211}\BibitemShut {NoStop}%
\bibitem [{\citenamefont {Krzakala}\ \emph {et~al.}(2013)\citenamefont
  {Krzakala}, \citenamefont {Moore}, \citenamefont {Mossel}, \citenamefont
  {Neeman}, \citenamefont {Sly}, \citenamefont {Zdeborov{\'a}},\ and\
  \citenamefont {Zhang}}]{krzakala2013spectral}%
  \BibitemOpen
  \bibfield  {author} {\bibinfo {author} {\bibfnamefont {F.}~\bibnamefont
  {Krzakala}}, \bibinfo {author} {\bibfnamefont {C.}~\bibnamefont {Moore}},
  \bibinfo {author} {\bibfnamefont {E.}~\bibnamefont {Mossel}}, \bibinfo
  {author} {\bibfnamefont {J.}~\bibnamefont {Neeman}}, \bibinfo {author}
  {\bibfnamefont {A.}~\bibnamefont {Sly}}, \bibinfo {author} {\bibfnamefont
  {L.}~\bibnamefont {Zdeborov{\'a}}}, \ and\ \bibinfo {author} {\bibfnamefont
  {P.}~\bibnamefont {Zhang}},\ }\bibfield  {title} {\enquote {\bibinfo {title}
  {Spectral redemption in clustering sparse networks},}\ }\href@noop {}
  {\bibfield  {journal} {\bibinfo  {journal} {Proc. Natl. Acad. Sci. USA}\
  }\textbf {\bibinfo {volume} {110}},\ \bibinfo {pages} {20935} (\bibinfo
  {year} {2013})}\BibitemShut {NoStop}%
\bibitem [{\citenamefont {Weiss}\ and\ \citenamefont
  {Freeman}(2001)}]{weiss2001optimality}%
  \BibitemOpen
  \bibfield  {author} {\bibinfo {author} {\bibfnamefont {Y.}~\bibnamefont
  {Weiss}}\ and\ \bibinfo {author} {\bibfnamefont {W.~T.}\ \bibnamefont
  {Freeman}},\ }\bibfield  {title} {\enquote {\bibinfo {title} {On the
  optimality of solutions of the max-product belief-propagation algorithm in
  arbitrary graphs},}\ }\href@noop {} {\bibfield  {journal} {\bibinfo
  {journal} {IEEE Trans. Info. Theory}\ }\textbf {\bibinfo {volume} {47}},\
  \bibinfo {pages} {736} (\bibinfo {year} {2001})}\BibitemShut {NoStop}%
\bibitem [{\citenamefont {Faqeeh}\ \emph {et~al.}(2015)\citenamefont {Faqeeh},
  \citenamefont {Melnik},\ and\ \citenamefont {Gleeson}}]{faqeeh2015network}%
  \BibitemOpen
  \bibfield  {author} {\bibinfo {author} {\bibfnamefont {A.}~\bibnamefont
  {Faqeeh}}, \bibinfo {author} {\bibfnamefont {S.}~\bibnamefont {Melnik}}, \
  and\ \bibinfo {author} {\bibfnamefont {J.~P.}\ \bibnamefont {Gleeson}},\
  }\bibfield  {title} {\enquote {\bibinfo {title} {Network cloning unfolds the
  effect of clustering on dynamical processes},}\ }\href@noop {} {\bibfield
  {journal} {\bibinfo  {journal} {Phys. Rev. E}\ }\textbf {\bibinfo {volume}
  {91}},\ \bibinfo {pages} {052807} (\bibinfo {year} {2015})}\BibitemShut
  {NoStop}%
\bibitem [{\citenamefont {Minc}(1988)}]{minc1988nonnegative}%
  \BibitemOpen
  \bibfield  {author} {\bibinfo {author} {\bibfnamefont {H.}~\bibnamefont
  {Minc}},\ }\href@noop {} {\emph {\bibinfo {title} {Nonnegative Matrices}}}\
  (\bibinfo  {publisher} {John Wiley and Sons, New York},\ \bibinfo {year}
  {1988})\BibitemShut {NoStop}%
\bibitem [{\citenamefont {Martin}\ \emph {et~al.}(2014)\citenamefont {Martin},
  \citenamefont {Zhang},\ and\ \citenamefont
  {Newman}}]{martin2014localization}%
  \BibitemOpen
  \bibfield  {author} {\bibinfo {author} {\bibfnamefont {T.}~\bibnamefont
  {Martin}}, \bibinfo {author} {\bibfnamefont {X.}~\bibnamefont {Zhang}}, \
  and\ \bibinfo {author} {\bibfnamefont {M.~E.~J.}\ \bibnamefont {Newman}},\
  }\bibfield  {title} {\enquote {\bibinfo {title} {Localization and centrality
  in networks},}\ }\href@noop {} {\bibfield  {journal} {\bibinfo  {journal}
  {Phys. Rev. E}\ }\textbf {\bibinfo {volume} {90}},\ \bibinfo {pages} {052808}
  (\bibinfo {year} {2014})}\BibitemShut {NoStop}%
\bibitem [{\citenamefont {Hamilton}\ and\ \citenamefont
  {Pryadko}(2014)}]{hamilton2014tight}%
  \BibitemOpen
  \bibfield  {author} {\bibinfo {author} {\bibfnamefont {K.~E.}\ \bibnamefont
  {Hamilton}}\ and\ \bibinfo {author} {\bibfnamefont {L.~P.}\ \bibnamefont
  {Pryadko}},\ }\bibfield  {title} {\enquote {\bibinfo {title} {Tight lower
  bound for percolation threshold on an infinite graph},}\ }\href@noop {}
  {\bibfield  {journal} {\bibinfo  {journal} {Phys. Rev. Lett.}\ }\textbf
  {\bibinfo {volume} {113}},\ \bibinfo {pages} {208701} (\bibinfo {year}
  {2014})}\BibitemShut {NoStop}%
\bibitem [{\citenamefont {Goltsev}\ \emph {et~al.}(2008)\citenamefont
  {Goltsev}, \citenamefont {Dorogovtsev},\ and\ \citenamefont
  {Mendes}}]{goltsev2008percolation}%
  \BibitemOpen
  \bibfield  {author} {\bibinfo {author} {\bibfnamefont {A.~V.}\ \bibnamefont
  {Goltsev}}, \bibinfo {author} {\bibfnamefont {S.~N.}\ \bibnamefont
  {Dorogovtsev}}, \ and\ \bibinfo {author} {\bibfnamefont {J.~F.~F.}\
  \bibnamefont {Mendes}},\ }\bibfield  {title} {\enquote {\bibinfo {title}
  {Percolation on correlated networks},}\ }\href@noop {} {\bibfield  {journal}
  {\bibinfo  {journal} {Phys. Rev. E}\ }\textbf {\bibinfo {volume} {78}},\
  \bibinfo {pages} {051105} (\bibinfo {year} {2008})}\BibitemShut {NoStop}%
\bibitem [{\citenamefont {Goltsev}\ \emph {et~al.}(2012)\citenamefont
  {Goltsev}, \citenamefont {Dorogovtsev}, \citenamefont {Oliveira},\ and\
  \citenamefont {Mendes}}]{goltsev2012localization}%
  \BibitemOpen
  \bibfield  {author} {\bibinfo {author} {\bibfnamefont {A.~V.}\ \bibnamefont
  {Goltsev}}, \bibinfo {author} {\bibfnamefont {S.~N.}\ \bibnamefont
  {Dorogovtsev}}, \bibinfo {author} {\bibfnamefont {J.~G.}\ \bibnamefont
  {Oliveira}}, \ and\ \bibinfo {author} {\bibfnamefont {J.~F.~F.}\ \bibnamefont
  {Mendes}},\ }\bibfield  {title} {\enquote {\bibinfo {title} {Localization and
  spreading of diseases in complex networks},}\ }\href@noop {} {\bibfield
  {journal} {\bibinfo  {journal} {Phys. Rev. Lett.}\ }\textbf {\bibinfo
  {volume} {109}},\ \bibinfo {pages} {128702} (\bibinfo {year}
  {2012})}\BibitemShut {NoStop}%
\bibitem [{\citenamefont {Saade}\ \emph {et~al.}(2014)\citenamefont {Saade},
  \citenamefont {Krzakala},\ and\ \citenamefont
  {Zdeborov{\'a}}}]{saade2014spectral}%
  \BibitemOpen
  \bibfield  {author} {\bibinfo {author} {\bibfnamefont {A.}~\bibnamefont
  {Saade}}, \bibinfo {author} {\bibfnamefont {F.}~\bibnamefont {Krzakala}}, \
  and\ \bibinfo {author} {\bibfnamefont {L.}~\bibnamefont {Zdeborov{\'a}}},\
  }\bibfield  {title} {\enquote {\bibinfo {title} {Spectral clustering of
  graphs with the {B}ethe {H}essian},}\ }in\ \href@noop {} {\emph {\bibinfo
  {booktitle} {Adv. Neural Inf. Process. Syst.}}}\ (\bibinfo {year} {2014})\
  p.\ \bibinfo {pages} {406}\BibitemShut {NoStop}%
\bibitem [{\citenamefont {Zachary}(1977)}]{zachary1977information}%
  \BibitemOpen
  \bibfield  {author} {\bibinfo {author} {\bibfnamefont {W.~W.}\ \bibnamefont
  {Zachary}},\ }\bibfield  {title} {\enquote {\bibinfo {title} {An information
  flow model for conflict and fission in small groups},}\ }\href@noop {}
  {\bibfield  {journal} {\bibinfo  {journal} {J. Anthropol. Res.}\ }\textbf
  {\bibinfo {volume} {33}},\ \bibinfo {pages} {452} (\bibinfo {year}
  {1977})}\BibitemShut {NoStop}%
\bibitem [{\citenamefont {Ripeanu}\ \emph {et~al.}(2002)\citenamefont
  {Ripeanu}, \citenamefont {Iamnitchi},\ and\ \citenamefont
  {Foster}}]{network_gnutella}%
  \BibitemOpen
  \bibfield  {author} {\bibinfo {author} {\bibfnamefont {M.}~\bibnamefont
  {Ripeanu}}, \bibinfo {author} {\bibfnamefont {A.}~\bibnamefont {Iamnitchi}},
  \ and\ \bibinfo {author} {\bibfnamefont {I.}~\bibnamefont {Foster}},\
  }\bibfield  {title} {\enquote {\bibinfo {title} {Mapping the {G}nutella
  network},}\ }\href@noop {} {\bibfield  {journal} {\bibinfo  {journal} {IEEE
  Internet Comput.}\ }\textbf {\bibinfo {volume} {6}},\ \bibinfo {pages} {50}
  (\bibinfo {year} {2002})}\BibitemShut {NoStop}%
\bibitem [{\citenamefont {Meyer}(2001)}]{meyer2001university}%
  \BibitemOpen
  \bibfield  {author} {\bibinfo {author} {\bibfnamefont {D.}~\bibnamefont
  {Meyer}},\ }\href@noop {} {\enquote {\bibinfo {title} {{University of Oregon
  Route Views Archive Project}},}\ } (\bibinfo {year} {2001})\BibitemShut
  {NoStop}%
\bibitem [{\citenamefont {Braunstein}\ \emph {et~al.}(2016)\citenamefont
  {Braunstein}, \citenamefont {Dall’Asta}, \citenamefont {Semerjian},\ and\
  \citenamefont {Zdeborov{\'a}}}]{braunstein2016network}%
  \BibitemOpen
  \bibfield  {author} {\bibinfo {author} {\bibfnamefont {A.}~\bibnamefont
  {Braunstein}}, \bibinfo {author} {\bibfnamefont {L.}~\bibnamefont
  {Dall’Asta}}, \bibinfo {author} {\bibfnamefont {G.}~\bibnamefont
  {Semerjian}}, \ and\ \bibinfo {author} {\bibfnamefont {L.}~\bibnamefont
  {Zdeborov{\'a}}},\ }\bibfield  {title} {\enquote {\bibinfo {title} {Network
  dismantling},}\ }\href@noop {} {\bibfield  {journal} {\bibinfo  {journal}
  {Proc. Natl. Acad. Sci. USA}\ }\textbf {\bibinfo {volume} {113}},\ \bibinfo
  {pages} {12368} (\bibinfo {year} {2016})}\BibitemShut {NoStop}%
\bibitem [{\citenamefont {Zdeborov{\'a}}\ \emph {et~al.}(2016)\citenamefont
  {Zdeborov{\'a}}, \citenamefont {Zhang},\ and\ \citenamefont
  {Zhou}}]{zdeborova2016fast}%
  \BibitemOpen
  \bibfield  {author} {\bibinfo {author} {\bibfnamefont {L.}~\bibnamefont
  {Zdeborov{\'a}}}, \bibinfo {author} {\bibfnamefont {P.}~\bibnamefont
  {Zhang}}, \ and\ \bibinfo {author} {\bibfnamefont {H.~J.}\ \bibnamefont
  {Zhou}},\ }\bibfield  {title} {\enquote {\bibinfo {title} {Fast and simple
  decycling and dismantling of networks},}\ }\href@noop {} {\bibfield
  {journal} {\bibinfo  {journal} {Sci. Rep.}\ }\textbf {\bibinfo {volume}
  {6}},\ \bibinfo {pages} {37954} (\bibinfo {year} {2016})}\BibitemShut
  {NoStop}%
\bibitem [{\citenamefont {Newman}\ and\ \citenamefont
  {Girvan}(2004)}]{newman2004finding}%
  \BibitemOpen
  \bibfield  {author} {\bibinfo {author} {\bibfnamefont {M.~E.~J.}\
  \bibnamefont {Newman}}\ and\ \bibinfo {author} {\bibfnamefont
  {M.}~\bibnamefont {Girvan}},\ }\bibfield  {title} {\enquote {\bibinfo {title}
  {Finding and evaluating community structure in networks},}\ }\href@noop {}
  {\bibfield  {journal} {\bibinfo  {journal} {Phys. Rev. E}\ }\textbf {\bibinfo
  {volume} {69}},\ \bibinfo {pages} {026113} (\bibinfo {year}
  {2004})}\BibitemShut {NoStop}%
\bibitem [{\citenamefont {Leskovec}\ \emph {et~al.}(2009)\citenamefont
  {Leskovec}, \citenamefont {Lang}, \citenamefont {Dasgupta},\ and\
  \citenamefont {Mahoney}}]{network_roads}%
  \BibitemOpen
  \bibfield  {author} {\bibinfo {author} {\bibfnamefont {J.}~\bibnamefont
  {Leskovec}}, \bibinfo {author} {\bibfnamefont {K.~J.}\ \bibnamefont {Lang}},
  \bibinfo {author} {\bibfnamefont {A.}~\bibnamefont {Dasgupta}}, \ and\
  \bibinfo {author} {\bibfnamefont {M.~W.}\ \bibnamefont {Mahoney}},\
  }\bibfield  {title} {\enquote {\bibinfo {title} {Community structure in large
  networks: Natural cluster sizes and the absence of large well-defined
  clusters},}\ }\href@noop {} {\bibfield  {journal} {\bibinfo  {journal}
  {Internet Math.}\ }\textbf {\bibinfo {volume} {6}},\ \bibinfo {pages} {29}
  (\bibinfo {year} {2009})}\BibitemShut {NoStop}%
\bibitem [{\citenamefont {Yang}\ and\ \citenamefont
  {Leskovec}(2015)}]{network_amazon}%
  \BibitemOpen
  \bibfield  {author} {\bibinfo {author} {\bibfnamefont {J.}~\bibnamefont
  {Yang}}\ and\ \bibinfo {author} {\bibfnamefont {J.}~\bibnamefont
  {Leskovec}},\ }\bibfield  {title} {\enquote {\bibinfo {title} {Defining and
  evaluating network communities based on ground-truth},}\ }\href@noop {}
  {\bibfield  {journal} {\bibinfo  {journal} {Knowl. Inf. Syst.}\ }\textbf
  {\bibinfo {volume} {42}},\ \bibinfo {pages} {181} (\bibinfo {year}
  {2015})}\BibitemShut {NoStop}%
\bibitem [{\citenamefont {Radicchi}\ and\ \citenamefont
  {Castellano}(2016)}]{radicchi2016beyond}%
  \BibitemOpen
  \bibfield  {author} {\bibinfo {author} {\bibfnamefont {F.}~\bibnamefont
  {Radicchi}}\ and\ \bibinfo {author} {\bibfnamefont {C.}~\bibnamefont
  {Castellano}},\ }\bibfield  {title} {\enquote {\bibinfo {title} {Beyond the
  locally treelike approximation for percolation on real networks},}\
  }\href@noop {} {\bibfield  {journal} {\bibinfo  {journal} {Phys. Rev. E}\
  }\textbf {\bibinfo {volume} {93}},\ \bibinfo {pages} {030302} (\bibinfo
  {year} {2016})}\BibitemShut {NoStop}%
\bibitem [{\citenamefont {Yoon}\ \emph {et~al.}(2011)\citenamefont {Yoon},
  \citenamefont {Goltsev}, \citenamefont {Dorogovtsev},\ and\ \citenamefont
  {Mendes}}]{yoon2011belief}%
  \BibitemOpen
  \bibfield  {author} {\bibinfo {author} {\bibfnamefont {S.}~\bibnamefont
  {Yoon}}, \bibinfo {author} {\bibfnamefont {A.~V.}\ \bibnamefont {Goltsev}},
  \bibinfo {author} {\bibfnamefont {S.~N.}\ \bibnamefont {Dorogovtsev}}, \ and\
  \bibinfo {author} {\bibfnamefont {J.~F.~F.}\ \bibnamefont {Mendes}},\
  }\bibfield  {title} {\enquote {\bibinfo {title} {Belief-propagation algorithm
  and the {Ising} model on networks with arbitrary distributions of motifs},}\
  }\href@noop {} {\bibfield  {journal} {\bibinfo  {journal} {Phys. Rev. E}\
  }\textbf {\bibinfo {volume} {84}},\ \bibinfo {pages} {041144} (\bibinfo
  {year} {2011})}\BibitemShut {NoStop}%
\bibitem [{\citenamefont {Faqeeh}\ \emph {et~al.}(2016)\citenamefont {Faqeeh},
  \citenamefont {Melnik}, \citenamefont {Colomer-de Sim\'on},\ and\
  \citenamefont {Gleeson}}]{faqeeh2015emergence}%
  \BibitemOpen
  \bibfield  {author} {\bibinfo {author} {\bibfnamefont {A.}~\bibnamefont
  {Faqeeh}}, \bibinfo {author} {\bibfnamefont {S.}~\bibnamefont {Melnik}},
  \bibinfo {author} {\bibfnamefont {P.}~\bibnamefont {Colomer-de Sim\'on}}, \
  and\ \bibinfo {author} {\bibfnamefont {J.~P.}\ \bibnamefont {Gleeson}},\
  }\bibfield  {title} {\enquote {\bibinfo {title} {Emergence of coexisting
  percolating clusters in networks},}\ }\href@noop {} {\bibfield  {journal}
  {\bibinfo  {journal} {Phys. Rev. E}\ }\textbf {\bibinfo {volume} {93}},\
  \bibinfo {pages} {062308} (\bibinfo {year} {2016})}\BibitemShut {NoStop}%
\bibitem [{\citenamefont {Schaeffer}(2007)}]{schaeffer2007graph}%
  \BibitemOpen
  \bibfield  {author} {\bibinfo {author} {\bibfnamefont {S.~E.}\ \bibnamefont
  {Schaeffer}},\ }\bibfield  {title} {\enquote {\bibinfo {title} {Graph
  clustering},}\ }\href@noop {} {\bibfield  {journal} {\bibinfo  {journal}
  {Comput. Sci. Rev.}\ }\textbf {\bibinfo {volume} {1}},\ \bibinfo {pages} {27}
  (\bibinfo {year} {2007})}\BibitemShut {NoStop}%
\bibitem [{\citenamefont {Jarrell}\ \emph {et~al.}(2012)\citenamefont
  {Jarrell}, \citenamefont {Wang}, \citenamefont {Bloniarz}, \citenamefont
  {Brittin}, \citenamefont {Xu}, \citenamefont {Thomson}, \citenamefont
  {Albertson}, \citenamefont {Hall},\ and\ \citenamefont
  {Emmons}}]{jarrell2012connectome}%
  \BibitemOpen
  \bibfield  {author} {\bibinfo {author} {\bibfnamefont {T.~A.}\ \bibnamefont
  {Jarrell}}, \bibinfo {author} {\bibfnamefont {Y.}~\bibnamefont {Wang}},
  \bibinfo {author} {\bibfnamefont {A.~E.}\ \bibnamefont {Bloniarz}}, \bibinfo
  {author} {\bibfnamefont {C.~A.}\ \bibnamefont {Brittin}}, \bibinfo {author}
  {\bibfnamefont {M.}~\bibnamefont {Xu}}, \bibinfo {author} {\bibfnamefont
  {J.~N.}\ \bibnamefont {Thomson}}, \bibinfo {author} {\bibfnamefont {D.~G.}\
  \bibnamefont {Albertson}}, \bibinfo {author} {\bibfnamefont {D.~H.}\
  \bibnamefont {Hall}}, \ and\ \bibinfo {author} {\bibfnamefont {S.~W.}\
  \bibnamefont {Emmons}},\ }\bibfield  {title} {\enquote {\bibinfo {title} {The
  connectome of a decision-making neural network},}\ }\href@noop {} {\bibfield
  {journal} {\bibinfo  {journal} {Science}\ }\textbf {\bibinfo {volume}
  {337}},\ \bibinfo {pages} {437} (\bibinfo {year} {2012})}\BibitemShut
  {NoStop}%
\bibitem [{\citenamefont {Cohen}\ and\ \citenamefont
  {Havlin}(2003)}]{cohen2003scale}%
  \BibitemOpen
  \bibfield  {author} {\bibinfo {author} {\bibfnamefont {R.}~\bibnamefont
  {Cohen}}\ and\ \bibinfo {author} {\bibfnamefont {S.}~\bibnamefont {Havlin}},\
  }\bibfield  {title} {\enquote {\bibinfo {title} {Scale-free networks are
  ultrasmall},}\ }\href@noop {} {\bibfield  {journal} {\bibinfo  {journal}
  {Phys. Rev. Lett.}\ }\textbf {\bibinfo {volume} {90}},\ \bibinfo {pages}
  {058701} (\bibinfo {year} {2003})}\BibitemShut {NoStop}%
\bibitem [{\citenamefont {Dorogovtsev}\ \emph {et~al.}(2003)\citenamefont
  {Dorogovtsev}, \citenamefont {Mendes},\ and\ \citenamefont
  {Samukhin}}]{dorogovtsev2003metric}%
  \BibitemOpen
  \bibfield  {author} {\bibinfo {author} {\bibfnamefont {S.~N.}\ \bibnamefont
  {Dorogovtsev}}, \bibinfo {author} {\bibfnamefont {J.~F.~F.}\ \bibnamefont
  {Mendes}}, \ and\ \bibinfo {author} {\bibfnamefont {A.~N.}\ \bibnamefont
  {Samukhin}},\ }\bibfield  {title} {\enquote {\bibinfo {title} {Metric
  structure of random networks},}\ }\href@noop {} {\bibfield  {journal}
  {\bibinfo  {journal} {Nucl. Phys. B}\ }\textbf {\bibinfo {volume} {653}},\
  \bibinfo {pages} {307} (\bibinfo {year} {2003})}\BibitemShut {NoStop}%
\bibitem [{\citenamefont {Buldyrev}\ \emph {et~al.}(2010)\citenamefont
  {Buldyrev}, \citenamefont {Parshani}, \citenamefont {Paul}, \citenamefont
  {Stanley},\ and\ \citenamefont {Havlin}}]{buldyrev2010catastrophic}%
  \BibitemOpen
  \bibfield  {author} {\bibinfo {author} {\bibfnamefont {S.~V.}\ \bibnamefont
  {Buldyrev}}, \bibinfo {author} {\bibfnamefont {R.}~\bibnamefont {Parshani}},
  \bibinfo {author} {\bibfnamefont {G.}~\bibnamefont {Paul}}, \bibinfo {author}
  {\bibfnamefont {H.~E.}\ \bibnamefont {Stanley}}, \ and\ \bibinfo {author}
  {\bibfnamefont {S.}~\bibnamefont {Havlin}},\ }\bibfield  {title} {\enquote
  {\bibinfo {title} {Catastrophic cascade of failures in interdependent
  networks},}\ }\href@noop {} {\bibfield  {journal} {\bibinfo  {journal}
  {Nature}\ }\textbf {\bibinfo {volume} {464}},\ \bibinfo {pages} {1025}
  (\bibinfo {year} {2010})}\BibitemShut {NoStop}%
\bibitem [{\citenamefont {Bianconi}\ and\ \citenamefont
  {Dorogovtsev}(2014)}]{bianconi2014multiple}%
  \BibitemOpen
  \bibfield  {author} {\bibinfo {author} {\bibfnamefont {G.}~\bibnamefont
  {Bianconi}}\ and\ \bibinfo {author} {\bibfnamefont {S.~N.}\ \bibnamefont
  {Dorogovtsev}},\ }\bibfield  {title} {\enquote {\bibinfo {title} {Multiple
  percolation transitions in a configuration model of a network of networks},}\
  }\href@noop {} {\bibfield  {journal} {\bibinfo  {journal} {Phys. Rev. E}\
  }\textbf {\bibinfo {volume} {89}},\ \bibinfo {pages} {062814} (\bibinfo
  {year} {2014})}\BibitemShut {NoStop}%
\bibitem [{\citenamefont {Boccaletti}\ \emph {et~al.}(2014)\citenamefont
  {Boccaletti}, \citenamefont {Bianconi}, \citenamefont {Criado}, \citenamefont
  {del Genio}, \citenamefont {G{\'o}mez-Garde\~{n}es}, \citenamefont {Romance},
  \citenamefont {{Sendi\~{n}a-Nadal}}, \citenamefont {Wang},\ and\
  \citenamefont {Zanin}}]{boccaletti2014structure}%
  \BibitemOpen
  \bibfield  {author} {\bibinfo {author} {\bibfnamefont {S.}~\bibnamefont
  {Boccaletti}}, \bibinfo {author} {\bibfnamefont {G.}~\bibnamefont
  {Bianconi}}, \bibinfo {author} {\bibfnamefont {R.}~\bibnamefont {Criado}},
  \bibinfo {author} {\bibfnamefont {C.~I.}\ \bibnamefont {del Genio}}, \bibinfo
  {author} {\bibfnamefont {J.}~\bibnamefont {G{\'o}mez-Garde\~{n}es}}, \bibinfo
  {author} {\bibfnamefont {M.}~\bibnamefont {Romance}}, \bibinfo {author}
  {\bibfnamefont {I.}~\bibnamefont {{Sendi\~{n}a-Nadal}}}, \bibinfo {author}
  {\bibfnamefont {Z.}~\bibnamefont {Wang}}, \ and\ \bibinfo {author}
  {\bibfnamefont {M.}~\bibnamefont {Zanin}},\ }\bibfield  {title} {\enquote
  {\bibinfo {title} {The structure and dynamics of multilayer networks},}\
  }\href@noop {} {\bibfield  {journal} {\bibinfo  {journal} {Phys. Rep.}\
  }\textbf {\bibinfo {volume} {544}},\ \bibinfo {pages} {1} (\bibinfo {year}
  {2014})}\BibitemShut {NoStop}%
\bibitem [{\citenamefont {Kivel{\"a}}\ \emph {et~al.}(2014)\citenamefont
  {Kivel{\"a}}, \citenamefont {Arenas}, \citenamefont {Barthelemy},
  \citenamefont {Gleeson}, \citenamefont {Moreno},\ and\ \citenamefont
  {Porter}}]{kivela2014multilayer}%
  \BibitemOpen
  \bibfield  {author} {\bibinfo {author} {\bibfnamefont {M.}~\bibnamefont
  {Kivel{\"a}}}, \bibinfo {author} {\bibfnamefont {A.}~\bibnamefont {Arenas}},
  \bibinfo {author} {\bibfnamefont {M.}~\bibnamefont {Barthelemy}}, \bibinfo
  {author} {\bibfnamefont {J.~P.}\ \bibnamefont {Gleeson}}, \bibinfo {author}
  {\bibfnamefont {Y.}~\bibnamefont {Moreno}}, \ and\ \bibinfo {author}
  {\bibfnamefont {M.~A.}\ \bibnamefont {Porter}},\ }\bibfield  {title}
  {\enquote {\bibinfo {title} {Multilayer networks},}\ }\href@noop {}
  {\bibfield  {journal} {\bibinfo  {journal} {J. Complex Networks}\ }\textbf
  {\bibinfo {volume} {2}},\ \bibinfo {pages} {203} (\bibinfo {year}
  {2014})}\BibitemShut {NoStop}%
\bibitem [{\citenamefont {Radicchi}(2015)}]{radicchi2015percolation}%
  \BibitemOpen
  \bibfield  {author} {\bibinfo {author} {\bibfnamefont {F.}~\bibnamefont
  {Radicchi}},\ }\bibfield  {title} {\enquote {\bibinfo {title} {Percolation in
  real interdependent networks},}\ }\href@noop {} {\bibfield  {journal}
  {\bibinfo  {journal} {Nature Phys.}\ }\textbf {\bibinfo {volume} {11}},\
  \bibinfo {pages} {597} (\bibinfo {year} {2015})}\BibitemShut {NoStop}%
\bibitem [{\citenamefont {Radicchi}\ and\ \citenamefont
  {Bianconi}(2017)}]{radicchi2017redundant}%
  \BibitemOpen
  \bibfield  {author} {\bibinfo {author} {\bibfnamefont {F.}~\bibnamefont
  {Radicchi}}\ and\ \bibinfo {author} {\bibfnamefont {G.}~\bibnamefont
  {Bianconi}},\ }\bibfield  {title} {\enquote {\bibinfo {title} {Redundant
  interdependencies boost the robustness of multiplex networks},}\ }\href@noop
  {} {\bibfield  {journal} {\bibinfo  {journal} {Phys. Rev. X}\ }\textbf
  {\bibinfo {volume} {7}},\ \bibinfo {pages} {011013} (\bibinfo {year}
  {2017})}\BibitemShut {NoStop}%
\bibitem [{\citenamefont {Baxter}\ \emph {et~al.}(2016)\citenamefont {Baxter},
  \citenamefont {Bianconi}, \citenamefont {da~Costa}, \citenamefont
  {Dorogovtsev},\ and\ \citenamefont {Mendes}}]{baxter2016correlated}%
  \BibitemOpen
  \bibfield  {author} {\bibinfo {author} {\bibfnamefont {G.~J.}\ \bibnamefont
  {Baxter}}, \bibinfo {author} {\bibfnamefont {G.}~\bibnamefont {Bianconi}},
  \bibinfo {author} {\bibfnamefont {R.~A.}\ \bibnamefont {da~Costa}}, \bibinfo
  {author} {\bibfnamefont {S.~N.}\ \bibnamefont {Dorogovtsev}}, \ and\ \bibinfo
  {author} {\bibfnamefont {J.~F.~F.}\ \bibnamefont {Mendes}},\ }\bibfield
  {title} {\enquote {\bibinfo {title} {Correlated edge overlaps in multiplex
  networks},}\ }\href@noop {} {\bibfield  {journal} {\bibinfo  {journal} {Phys.
  Rev. E}\ }\textbf {\bibinfo {volume} {94}},\ \bibinfo {pages} {012303}
  (\bibinfo {year} {2016})}\BibitemShut {NoStop}%
\end{thebibliography}
%

\end{document}